\documentclass[aps,prb,twocolumn,superscriptaddress,showpacs,amsmath,amssymb, amsfonts,10pt]{revtex4-1}
\usepackage[T1]{fontenc}
\usepackage[utf8]{inputenc}
\usepackage{amsmath,bm}
\usepackage{amssymb}
\usepackage{yfonts}
\usepackage{eufrak}
\usepackage{bbm}
\usepackage{braket}
\usepackage{graphicx}
\usepackage{xcolor}
\usepackage{pifont}
\usepackage[mathscr]{euscript}
\usepackage[shortlabels]{enumitem}
\usepackage{float}
\usepackage{dsfont}
\usepackage{comment}
\usepackage{slashed}

\usepackage[normalem]{ulem} 
\usepackage{bbm} 
\usepackage[percent]{overpic}

\usepackage[colorlinks=true]{hyperref}  
\hypersetup{
	unicode=false,          
	pdftoolbar=true,        
	pdfmenubar=true,        
	pdffitwindow=false,     
	pdfstartview={FitH},    
	pdftitle={FloquetAncilla},    
	pdfauthor={},     
	pdfsubject={},   
	pdfcreator={},   
	pdfproducer={}, 
	pdfkeywords={} {} {}, 
	pdfnewwindow=true,      
	colorlinks=true,       
	linkcolor=blue, 
	citecolor=blue,        
	filecolor=magenta,      
	urlcolor=blue           
} 

\newcommand{\change}[1]{{\color{black}#1}}

\usepackage{xcolor}
\definecolor{brandeisblue}{rgb}{0.0, 0.44, 1.0}

\usepackage{mathtools}

\newcommand{\bq}{\mathbf{q}}
\newcommand{\cl}{\mathrm{cl}}
\newcommand{\qn}{\mathrm{q}}
\newcommand{\rr}{\mathbf{r}}

\newcommand{\qq}{\mathbf{q}}
\newcommand{\bQ}{\mathbf{Q}}

\setcitestyle{numbers,square}      

\begin{document}
	
\title{Steady-states and response functions of the\\ periodically driven O($N$) scalar field theory}

\begin{abstract} 
	We investigate the phase diagram of a relativistic, parametrically driven O($N$)-symmetric theory coupled to a Markovian thermal bath. Our analysis reveals a rich variety of phases, including both uniform and spatially modulated symmetry-broken states, some of which feature an order parameter oscillating at half the drive frequency. When coupled to a background electromagnetic potential, these phases exhibit a Meissner effect, in the sense that the photon acquires a mass term. However, if the order parameter oscillates around a sufficiently small value, a fraction of an externally applied magnetic field can penetrate the sample in the form of a standing wave. We dub this property a \textit{Meissner polariton}, that is, a collective mode resulting from the hybridization of light with order parameter oscillations. 
	Furthermore, near the onset of symmetry breaking, strong fluctuations give rise to a superconducting-like response even in the absence of a Meissner effect or of a Meissner polariton. Our results are relevant to experiments on light-induced orders, particularly superconductivity.
\end{abstract}

\author{Oriana K. Diessel}
\affiliation{ITAMP, Harvard-Smithsonian Center for Astrophysics, Cambridge, MA 02138, USA}
\affiliation{Department of Physics, Harvard University, Cambridge MA 02138, USA}

\author{Subir Sachdev}
\email{Corresponding author: sachdev@g.harvard.edu}
\affiliation{Department of Physics, Harvard University, Cambridge MA 02138, USA}
\affiliation{Center for Computational Quantum Physics, Flatiron Institute, 162 5th Avenue, New York, NY 10010, USA}

\author{Pietro M. Bonetti}
\affiliation{Department of Physics, Harvard University, Cambridge MA 02138, USA}
\affiliation{Max Planck Institute for Solid State Research, Heisenbergstraße 1, D-70569 Stuttgart, Germany}

\maketitle
\newpage
\linespread{1.05}
\section{Introduction}

The past decade has seen remarkable progress in realizing possible new light-induced phases of matter. Examples include superconductivity in K$_3$C$_{60}$ \cite{Mitrano16}, the cuprates \cite{Cremin19,Shimano23,Shimano24,Cavalleri24,Comin25,Michael2024,Michael2025}, and the nickelates \cite{Xu25}, charge density waves \cite{Gedik20,Gedik21,Carpene24,Gedik25a,Gedik25b,Padma2025}, antiferromagnetism \cite{Gedik24} and ferroelectricity~\cite{Nova2019,Fechner2024} \change{(we note that the experimental observation of light-induced superconducting phases has been recently challenged~\cite{Dodge23,Comment23,Dodge25})}.
Motivated by these results, we examine a general theory of order parameter dynamics in the presence of a time-periodic modulation of the coupling constants of the effective free energy of the order parameter. We assume that the periodic driving does not break any symmetry associated with the order parameter, and study the phase diagram in which the symmetry may be spontaneously broken in some phases. Our focus is on the regime where the natural frequency of order parameter dynamics is comparable to the driving frequency $\Omega$, and the damping from the environment is moderately weak. In such a regime, a Floquet point-of-view is appropriate at linear order, where we consider frequencies shifted by integer multiples of $\Omega$. However, we will also study the full non-linear case, where all frequencies couple. 
\begin{figure*}
	\centering
	\includegraphics[width=1\linewidth]{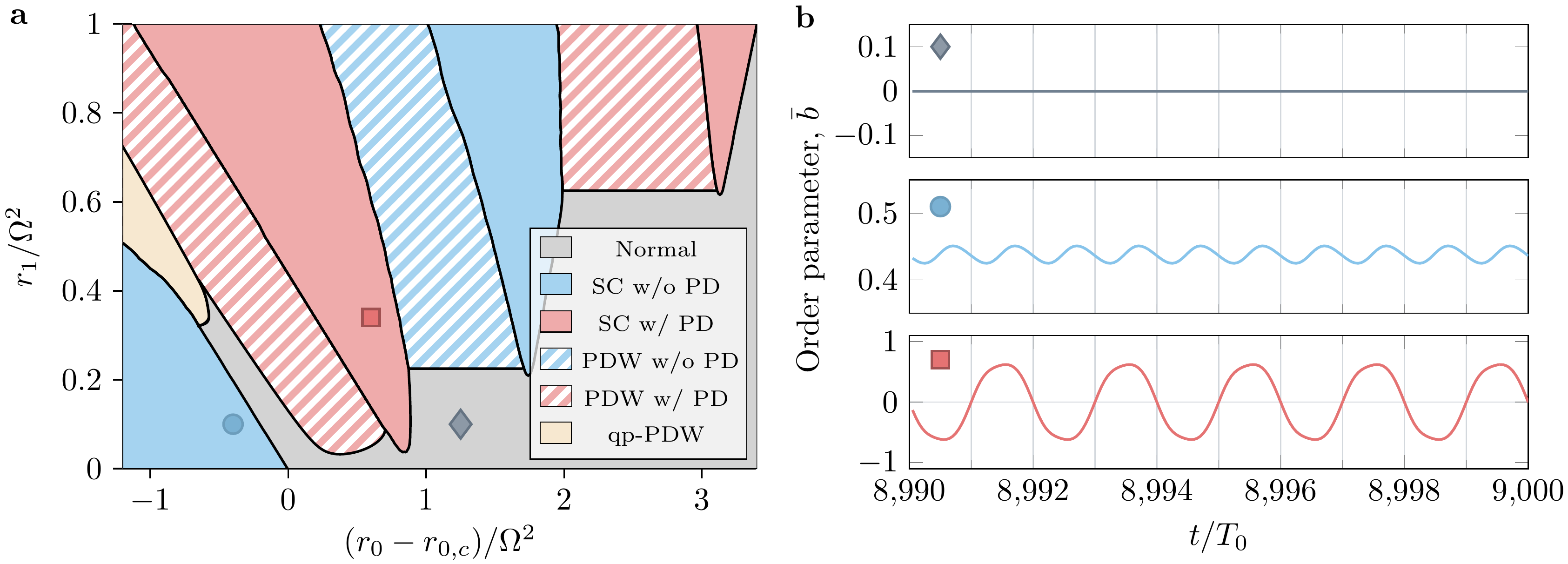}
	\caption{Panel (a): Phase diagram obtained by solving Eqs.~\eqref{eq:EoM} with parameters $u=\Omega^2$, and $\gamma=0.03\,\Omega$. Six phases appear: normal (gray), superconducting (SC,blue), period-doubling SC (red), pair density wave (PDW, blue hatches), PDW with period doubling (red hatches) and a quasi-periodic PDW (yellow). 
		The symbols mark the points whose order-parameter time traces over the last ten drive periods are shown in panel (b): \change{the circle at $(r_0-r_{0,c})/\Omega^2=-0.4$, $r_1/\Omega^2=0.1$, the square at $(0.6,0.34)$, and the diamond at $(1.25,0.1)$.} As detailed in App.~\ref{app: finite-q}, we also performed a dynamical stability analysis. The hatched regions indicate an instability of the normal state towards a PDW, oscillating at either half the drive frequency (red hatched regions), labeled as PDW with period doubling (PDW w PD), or a at the drive frequency, labeled as PDW without period doubling (PDW w/o PD). The yellow region indicates a portion of the phase diagram where the steady state is non-periodic but the correlation function $C_\bq(t)$ diverges at finite $\bq$. We labeled this region as quasi-periodic PDW (qp-PDW). Panel (b): Representative dynamics $\bar{b}(t)$ over the last ten periods, plotted versus $t/T_0$, for the parameter points indicated by the symbols in panel (a). Top: metal ($\bar{b}=0$), middle: SC (small oscillations with the drive period about a nonzero mean), bottom: period-doubling SC (oscillations with period $2T_0$ around zero).}
	\label{Fig:PhaseDiagram}
\end{figure*}

We will consider the simplest case of an O($N$) symmetric theory of a real $N$-component order parameter $b_\alpha$ with a 
quadratic, time-dependent `mass' term $r(t)$, and a
quartic, time-independent self-interaction $u$. The time dependence is specified by
\begin{equation}\label{eq: rt definition}
	r(t)=r_0+2 r_1\text{cos}(\Omega t)\,.
\end{equation}
so that $r_0$ is the tuning parameter across the equilibrium transition, and $r_1$ is the amplitude of the periodic driving. \change{A number of earlier studies have obtained related results on similar models \cite{Chandran16,Millis20,DaiLee21,Mitra21a,Mitra21b,Dai22,Sols24,Chandra25a,Moessner25,Chandra25b}.
	%
	%
	In particular, out-of-equilibrium, self-interacting O($N$) models have been considered in Refs.~\cite{Weidinger2017,Zelle2024,Daviet2024,Zelle25,ZelleMillis25}. It was shown that they exhibit a rich phenomenology, displaying, among others, prethermal and limit cycle phases.} \change{In Ref.~\cite{Homann2020}, resonant optical driving of a high-$T_c$ superconductor at the sum of its Higgs and Josephson plasma frequencies produces persistent subharmonic oscillations of the order parameter, identified as a Higgs time-crystalline phase.}

In this work, we calculate the phase diagram in the presence of \textit{both} self-interaction and dissipation. 
The phase diagram is shown in Fig.~\ref{Fig:PhaseDiagram}a as a function $r_0/\Omega^2$ and $r_1/\Omega^2$, and is obtained by
various analytic arguments, and a full numerical solution of the large $N$ equations. We use the terminology of the $N=2$ case which applies to the superconductor-normal transition. The long-time steady-state time-dependence of the O($N$) order parameter (SC) $b$ is shown in Fig.~\ref{Fig:PhaseDiagram}b. In the `SC' phase, the order parameter oscillates at a frequency $\Omega$ and 
non-zero mean, while in the `SC with PD' phase (sometimes called a `time crystal') there is a period-doubling and the order parameter oscillates at a frequency $\Omega/2$ with zero mean. Our dynamical stability analysis also unveils phases where the order parameter condenses at finite momentum, similar to previous discussions in the context of parametrically driven bosonic system~\cite{Dai2021,Chandra25a,Chandra25b,Kaplan2025b}.

Section~\ref{sec:em response} obtains results on response functions of the various phases of Fig.~\ref{Fig:PhaseDiagram}. For the $N=2$ case, the response function is the optical conductivity, associated with coupling of an external electromagnetic field to the conserved O($N$) current. We show that whenever the order parameter develops an instantaneous expectation value, a mass term for the photon appears. When the order parameter oscillates around a large offset, this leads to a conventional Meissner effect, with the magnetic field being fully expelled from the material. Conversely, when such offset is small or vanishing, a fraction of the external magnetic field can penetrate the sample in the form of a standing wave, a \textit{Meissner polariton}, that is, a collective mode emerging from coupling the order parameter oscillations to light. Moreover, we show that near the onset of a phase with an oscillating order parameter, the imaginary part of the optical conductivity, $\sigma_2(\omega)$, can be substantially enhanced and display a $1/\omega$ behavior over a broad interval of frequencies. This observation offers a possible explanation of the photo-induced enhancement of $\sigma_2(\omega)$ that has been experimentally observed. Previously, microscopic theories of driven electron-phonon systems have been proposed as a possible explanation of light-induced superconductivity~\cite{Komnik16,Knap2016,Babadi17,Kennes2017,Eckhardt2024}.

In Section~\ref{sec: conclusions}, we summarize our results and discuss possible experimental implications of our findings.

\section{Model and large-$N$ limit}

We consider a O($N$)-symmetric field theory for a real, $N$-component order parameter $b_{\alpha}$, where $\alpha=1,...,N$, described by the Lagrangian
\begin{equation}\label{eq: b potential}
	\begin{split}
		&\mathcal{L}[b]=\frac{1}{2}(\partial_\mu b)^2-\change{V[b]}\,,\\
		&\change{V[b]=\frac{r}{2}b^2+\frac{u}{4N}b^4}\,,
	\end{split}
\end{equation}
where $b^2\equiv \sum_{\alpha}b_{\alpha}^2$, and $u>0$.
We assume then that the parameter $r$, which in equilibrium describes the distance from the critical point, is periodically driven as in Eq.~\eqref{eq: rt definition}. Model~\eqref{eq: b potential} can describe the dynamics of several order parameters that are not associated with a conserved quantity, such as an antiferromagnetic order parameter for $N=3$. Even though a superconducting order parameter is usually the conjugate variable of a conserved quantity, that is the particle number, model~\eqref{eq: b potential} can also be employed for $N=2$ to describe the dynamics of the superconducting order parameter in the Bose-Hubbard model~\cite{Dai22} or in the attractive Fermi-Hubbard model near integer filling, where a particle-hole symmetry emerges. 


In order to avoid infinite heating due to the periodic drive, we assume that the order parameter is coupled to a Markovian thermal bath at temperature $T$. The equation of motion of the field $b_{\alpha}$ is then given by the Langevin equation
\begin{equation}
	\ddot{b}_{\alpha}+\gamma \dot{b}_{\alpha}+\frac{\partial V}{\partial b_{\alpha}}=\xi_{\alpha},
	\label{Eq:Langevin}
\end{equation}
where $\gamma$ quantifies dissipation resulting from coupling to the bath, $V$ represents the potential of the order parameter, and $\xi_{\alpha}$ is a zero-mean Gaussian white noise describing the thermal fluctuations from the bath with correlation given by $\langle \xi_{\alpha}(t,\mathbf{x}) \xi_{\beta}(t',\mathbf{x}')\rangle=2\gamma T \delta_{\alpha,\beta}\delta(t-t')\delta(\mathbf{x}-\mathbf{x}')$. 

In absence of a drive and for the $N=2$, model~\eqref{Eq:Langevin} describes the dynamics of the metal-to-superconducting transition~\cite{Tinkham2004}.
In this case, for $r\geq r_c$ the order parameter expectation value is zero, corresponding to a symmetric, normal phase, while for $r<r_c$, the order parameter acquires a finite expectation value corresponding to the symmetry-broken, superconducting phase.

Eq.~\eqref{Eq:Langevin} can be solved analytically only in the large-$N$ limit. To this end, one performs a Hubbard-Stratonovich transformation to decouple the quartic interaction term and then derive the saddle-point equations in the large-$N$ limit. \change{It is known~\cite{berges2004introduction} that two-loop processes, which are excluded in this limit, play an important role in redistributing momentum and energy, which will eventually help the system thermalize. However, we do not aim to describe intrinsic many-body thermalization here, but rather the emergence of dynamical instabilities and non-equilibrium steady states in a driven open system. As it will become clear in what follows, a large-$N$ approximation is sufficient to obtain these phenomena. Moreover, we model damping and relaxation effects phenomenologically by introducing a Markovian noise.}

The derivation \change{of the large-$N$ equations} is shown in App.~\ref{app:HST and SP}, and we report here only the final result.
We assume a spatially homogeneous condensate and we posit $\langle b \rangle = (\bar{b},0,\dots,0)$. The equation of motion for the order parameter and the correlation functions $C_\qq(t) = \langle b(\qq,t)^2\rangle-\bar{b}^2$ and $D_\qq(t)=\langle [\partial_t b(\qq,t)]^2\rangle-(\partial_t\bar{b})^2$ are given by:
\begin{subequations}
	\label{eq:EoM}
	\begin{align}
		&\left[\partial_t^2+\gamma\partial_t + \lambda(t)\right]\bar{b}(t)=0\,, \label{eq:EoM1}\\
		&\left[\partial_t^2+\gamma\partial_t+2(q^2 +\lambda(t))\right]C_\qq(t)=2D_\qq(t)\,,\label{eq:EoM2}\\
		&\left(\partial_t+2\gamma\right)D_\qq(t)+\left[q^2+\lambda(t)\right]\partial_t C_\qq(t)=2\gamma T\,, \label{eq:EoM3} 
	\end{align}
\end{subequations}
where $\lambda(t)$ is determined self-consistently as
\begin{equation}
	\lambda(t) = r(t) + u\left[\bar{b}^2(t) + \int_\qq C_\qq(t) \right],
	\label{Eq:Lambda}
\end{equation}
and $\int_\bq \equiv\int_{\Lambda_\mathrm{IR}}^{\Lambda_\mathrm{UV}}\frac{d^2\mathbf{q}}{(2\pi)^2}$. Notice that, since we are interested in two-dimensional systems, as they are more relevant for pump-and-probe experiments, we include an infrared cutoff to get around the Mermin-Wagner theorem and allow the system to undergo phase transitions.


We notice that similar equations were derived for an isolated quantum system ($\gamma=0$) in Refs~\cite{Chandran2016,Natseh2021_2}.
In this respect, here we extend their analysis to an open system and show that multiple steady states can be stabilized, depending on the strength of $r_1$. 

\section{Floquet Analysis}

Given the presence of a periodic drive and dissipation, we expect the system to reach a periodic steady state, i.e., a state where all observables are periodic, although not necessarily with the same period as the drive, as discussed further below.
By assuming that, in the periodic steady state, $\lambda(t)$ oscillates with the same period as the drive,  we observe that Eq.~\eqref{eq:EoM1} takes the form of a Mathieu-Hill equation~\cite{Mathieu,Hill}.
In this case,  nontrivial solutions can be found using Floquet's theorem~\cite{Floquet}, i.e., by working in the Fourier basis spanned by integer multiples of the drive frequency. The order parameter can then be written as    
\begin{equation}\label{eq: Floquet b}
	\bar{b}(t)=e^{-i\nu t}\sum_nb_n \,e^{-in\Omega t}\,,
\end{equation}
where $\sum_n\equiv\sum_{n=-\infty}^{+\infty}$, and $\nu\in(-\Omega/2,\Omega/2]$ is a parameter allowing for a different periodicity than the drive itself. \change{As we shall see in Sec.~\ref{sect: Full Num Solution}, the assumption of $\lambda(t)$ being time periodic with period $2\pi/\Omega$ applies to the vast majority of the phase diagram in Fig.~\ref{Fig:PhaseDiagram}.}
Inserting~\eqref{eq: Floquet b} into~\eqref{eq:EoM1} and expanding $\lambda(t)=\sum_n\lambda_n e^{-in\Omega t}$, we get
\begin{equation}
	\label{eq:order_parameter_equation}
	\mathcal{K}_{nm}(\nu)\, b_m=0
\end{equation}
where $\mathcal{K}_{nm}(\nu)$ is an infinite matrix in the Floquet basis given by:
\begin{multline}
	\label{eq:Kmn_matrix}
	\mathcal{K}_{nm}(\nu) = \\\left[(\nu+n\Omega)^2+i\gamma(\nu+n\Omega)\right]\delta_{nm}
	-\sum_\ell\lambda_\ell\, \delta_{n,m+\ell}.
\end{multline}
It is evident from the previous equation that values of $\lambda_n$ with $n\neq 0$ are responsible to couple the different Floquet modes.
Nontrivial solutions of Eq.~\eqref{eq:order_parameter_equation} are obtained by imposing $\mathrm{det}\,\mathcal{K}(\nu)=0$, which in turn imposes a condition on $\lambda_n$. In the simplifying assumption where $\lambda_{|n|>1}=0$, which is valid in the high-frequency limit where $\Omega$ is the largest energy scale, we recover the Mathieu equation and we can study its nontrivial solutions as functions of $\lambda_0$ and $|\lambda_1|$ (note that $\lambda_{-1}=\lambda_1^*$).
\begin{figure}[t]
	\centering
	\includegraphics[width=1\linewidth]{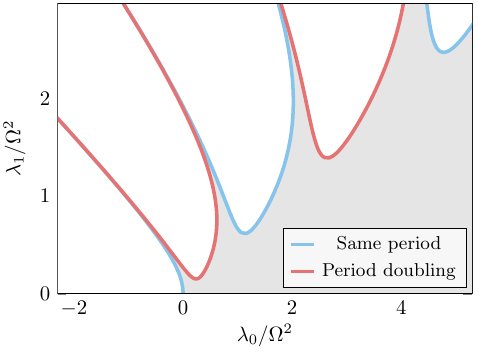}
	\caption{Lines in \change{$r_0$-$|r_1|$} plane where the damped Mathieu equation supports oscillating steady states at $\omega=0$ (blue lines) and at $\omega=\Omega/2$ (red lines) for $\gamma=0.3\Omega$. In the gray shaded regions, the solutions are decaying to zero, whereas in the unshaded regions solutions grow exponentially with time.}
	\label{fig:Arnold tongues}
\end{figure}

In Fig.~\ref{fig:Arnold tongues} we plot the lines in the \change{$r_0$-$|r_1|$} plane where $\mathrm{det}\,\mathcal{K}(\nu)=0$ for $\gamma/\Omega=0.3$ (solid blue and red lines). For any finite $\gamma$, $\mathcal{K}(\nu)$ has a vanishing eigenvalue only for $\nu=0,\Omega/2$.

The blue (red) lines in Fig.~\ref{fig:Arnold tongues} indicate the lines in parameter space where the Mathieu equation supports oscillating steady states with $\nu=0$ ($\nu=\Omega/2$).
The gray area indicates the parameter space where only solutions that decay to zero are supported. Conversely, in the white regions a parametric resonance occurs and $b(t)$ grows indefinitely with time. These regions are known as Arnold tongues in the mathematical literature~\cite{Arnold1965,Nayfeh1979,Kovacic2018,Strogatz2018}.

We can expand also Eq.~\eqref{Eq:Lambda} in drive harmonics in the Floquet basis for the two cases $\omega=0,\Omega/2$ as
\begin{equation}
	\label{eq: lambda equation floquet}
	\lambda_n = r_n + u \sum_m b_{n-m +\eta}b_m +u \int_\qq \!C_{n}(q)\,,
\end{equation}
with $\eta = 0$ for $\nu = 0$ and $\eta = 1$ for $\nu=\Omega/2$, $r_n$ is finite only for $n = -1, 0, 1$ since it is a pure sinusoidal drive, and $C_n(q)$ is the correlation function in the Floquet basis (see App.~\ref{app:C_Floquet}).

The self-consistent Floquet problem can be solved iteratively (see App.~\ref{app:truncation}) by truncating the number of Floquet modes to a large integer $N_\mathrm{fl}$, and provides an alternative way of solving Eqs~\eqref{eq:EoM} in the steady state. But even more importantly, it provides an analytical solution in some limit cases, as exemplified in the next section. 

\subsection{Analytical solutions in the high-frequency limit: Period-doubling solution}
\label{sec:Analytical solution}

In the following, we will be looking for period doubling ($\nu =\Omega/2$) steady states in the high-frequency limit, i.e., where $\Omega$ is the largest energy scale in the system.
To this end, we restrict ourselves to the Floquet modes $n=0$ and $n=-1$, which allows us to capture the solutions analytically. In order to capture the physics of the other Arnold tongues, a larger truncation in terms of Floquet modes is required.

\subsubsection{Mean-field solution}
\begin{figure}[t]
	\centering
	\includegraphics[width=1\linewidth]{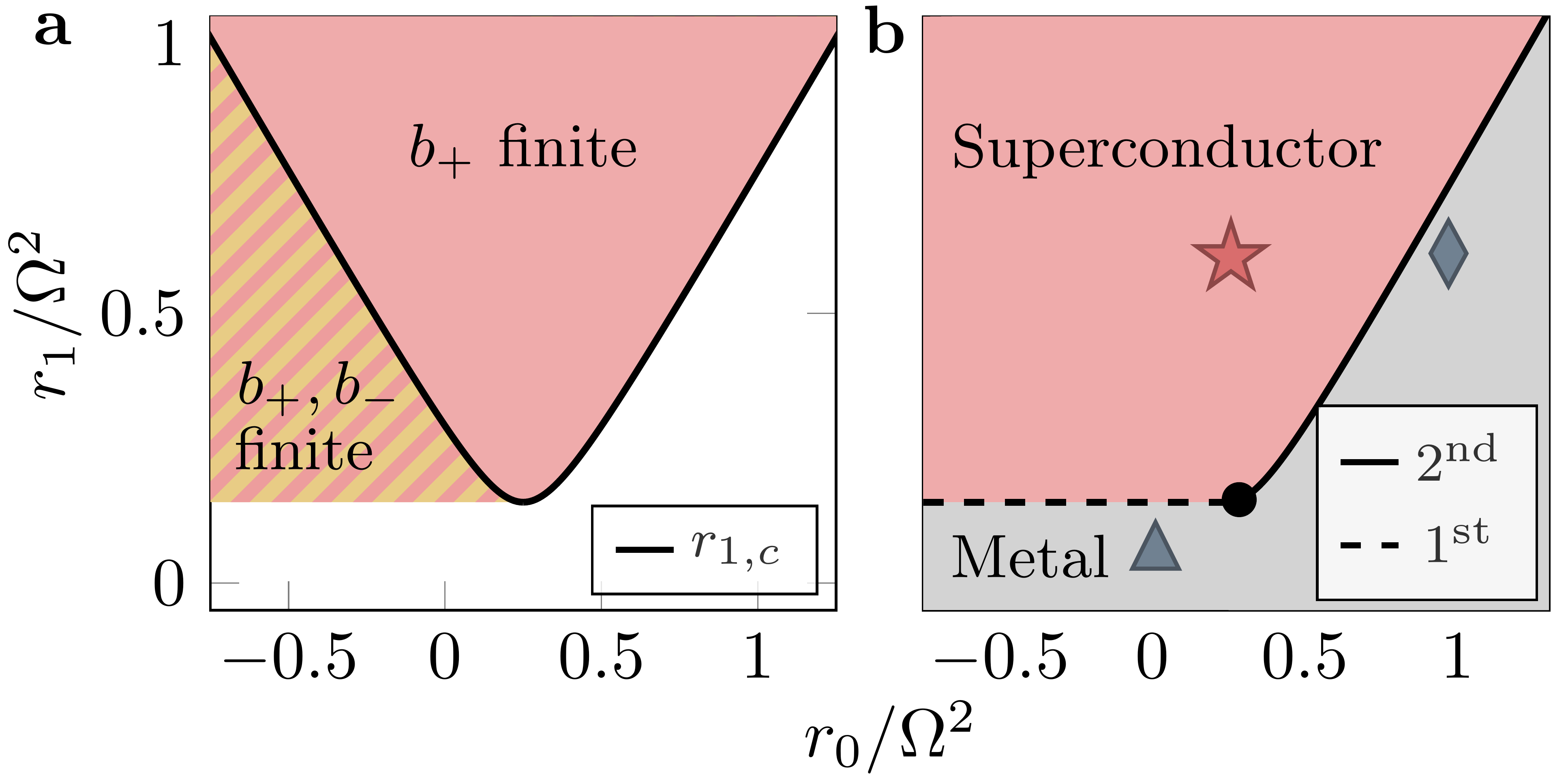}
	\caption{Panel (a): Location of the different finite mean-field solutions of the simplified two-modes model~\eqref{eq:K kernel analyitic} within the $r_0$-$r_1$ plane. The solid black curve denotes the curve~\eqref{eq:stability_condition}.  
		Panel (b): Phase diagram for the simplified two-mode model~\eqref{eq:K kernel analyitic}. The dashed line indicates a second-order phase transition, while the dashed one indicates a first order phase transitions. The large black dot indicates the exceptional point. The symbols (star, triangle, diamond) denotes points where we analyze the fluctuation dispersions (see Fig.~\ref{Fig:Dispersions}).
	}
	\label{Fig:OPMeanField}
\end{figure}

We now look for mean-field solutions, disregarding the effect of fluctuations.
We start from the case with $u=0$, where $\lambda_0 = r_0$ and $\lambda_1 = r_1$.
The condition $\mathrm{det}\mathcal{K}(\Omega/2)=0$ then reduces to:
\begin{equation}
	\label{eq:K kernel analyitic}
	0 = \text{det} \mathcal{K}\!\left(\frac{\Omega}{2}\right)\!\approx\det \!\left(\!
	\begin{array}{cc}
		\frac{\Omega^2}{4}-i\gamma \frac{\Omega}{2}-r_0 & -r_1 \\
		-r_1^* & \frac{\Omega^2}{4}+i\gamma \frac{\Omega}{2}-r_0
	\end{array}\!\right)\!,
\end{equation}
from which we can identify the curve in the $|r_1|-r_0$ plane where a finite and non-diverging solution of the order parameter exist, i.e.,
\begin{equation}
	\label{eq:stability_condition}
	|r_1| = r_{1,c}(r_0) \equiv \sqrt{\frac{\gamma^2 \Omega^2}{4}+ \left(r_0 - \frac{\Omega^2}{4}\right)^2}, 
\end{equation}
which correspond to the black solid line in Fig.~\ref{Fig:OPMeanField}, panel (a). For $|r_1| < r_{1,c}(r_0)$, the solutions are damped and therefore vanish, suggesting the system is in the symmetric (normal) phase, while for $|r_1| > r_{1,c}(r_0)$ the solution grows exponentially in time, which, while being unphysical, suggests the presence of an ordered, superconducting phase.

In order to make progress and stabilize the ordered phase we need to reinstate the nonlinearity $u$. We therefore insert $\nu = \Omega/2$ in Eq.~\eqref{eq: Floquet b} and retain only the $n=0,-1$ modes, so that the order parameter reads:
\begin{equation}
	\bar{b}(t) \approx b_0 e^{-i\frac{\Omega}{2}t} + b_{-1} e^{+i\frac{\Omega}{2}t}, 
\end{equation}
which entails $b_0^* = b_{-1}$, since $\bar{b}(t)$ is a real number. For simplicity we will define $b_0 \equiv b$ and $b_{-1} \equiv b^*$. Eq.~\eqref{eq: lambda equation floquet} then becomes:
\begin{equation}
	\lambda_0 \approx r_0 + 2u|b|^2, \qquad \lambda_1 \approx r_1. 
	\label{Eq:Lambda2}
\end{equation}
Condition~\eqref{eq:K kernel analyitic} is still valid, with $\lambda_0$ instead of $r_0$. Therefore, the stability condition~\eqref{eq:stability_condition} now becomes a condition to determine the value of the order parameter $|b|^2$. We found it admits two different solutions:
\begin{equation}
	|b_{\pm}|^2 = \frac{1}{2u}\left(\frac{\Omega^2}{4}-r_0 \pm\sqrt{|r_1|^2 -\frac{\gamma^2\Omega^2}{4}}\right)\,,
\end{equation}
which are defined up to a global phase, as expected for the SSB of a U(1) symmetric model.
$b_{+}$ and $b_{-}$ have finite values only for certain values of $r_0$ and $r_1$, as shown in Fig.~\ref{Fig:OPMeanField}, panel (a). $b_{+}$ is finite everywhere inside the hyperbole identified by Eq.~\eqref{eq:stability_condition}, while, remarkably, both $b_{\pm}$ are finite outside this region, where we first assumed the system would be in the normal phase.  
To shed more light on the phase diagram, we will study the dynamical stability of these solutions.

\subsubsection{Dynamical stability analysis}
\label{subsec: dynamical stability}
In order to study the dynamical stability of the different solutions (i.e., $b=0,b_\pm$) we study the poles of the truncated retarded Green functions evaluated in those solutions. Physically, this gives the spectrum of the excitations on top of those putative ground states (i.e., the normal or the superconducting state). Dynamical instabilities occur when the imaginary part of the excitation spectrum becomes positive for certain momenta $\mathbf{q}$, as it corresponds to an excitation indefinitely growing.
The truncated retarded Green's function reads:
\begin{multline}
	\label{eq:retarded}
	G^{-1}_R(\bq,\omega)\approx  \\
	\begin{pmatrix}
		(\omega\!-\!\Omega)^2+i\gamma(\omega\!-\!\Omega)-\omega_q^2 
		& -\lambda_1 \\
		-\lambda_1^* & \omega^2+i\gamma\omega-\omega_q^2 
	\end{pmatrix},
\end{multline}
with $\omega^2_q = \lambda_0+q^2$. Accordingly, the stability of the different solutions can be studied by replacing the corresponding value for the order parameter into Eq.~\eqref{Eq:Lambda2} and $\lambda_1 = r_1$.
The analysis of the poles of the retarded Green's function shows the following. First, the solution $b_{-}$ is never stable, and therefore it is unphysical. Second, the normal solution ($\bar{b}=0$) is unstable in the region to the left of the hyperbole (see Fig.~\ref{Fig:OPMeanField}), where the superconducting solution ($b_{+}$) is instead stable. The left branch of the hyperbole therefore does not represent the boundary between two phases.
Third, the dashed line at $r_1 = \gamma\Omega/2$ on the left of the tip of the parabola emerges as a boundary between the normal and the superconducting phase. 
\begin{figure}[t]
	\centering
	\includegraphics[width=1\linewidth]{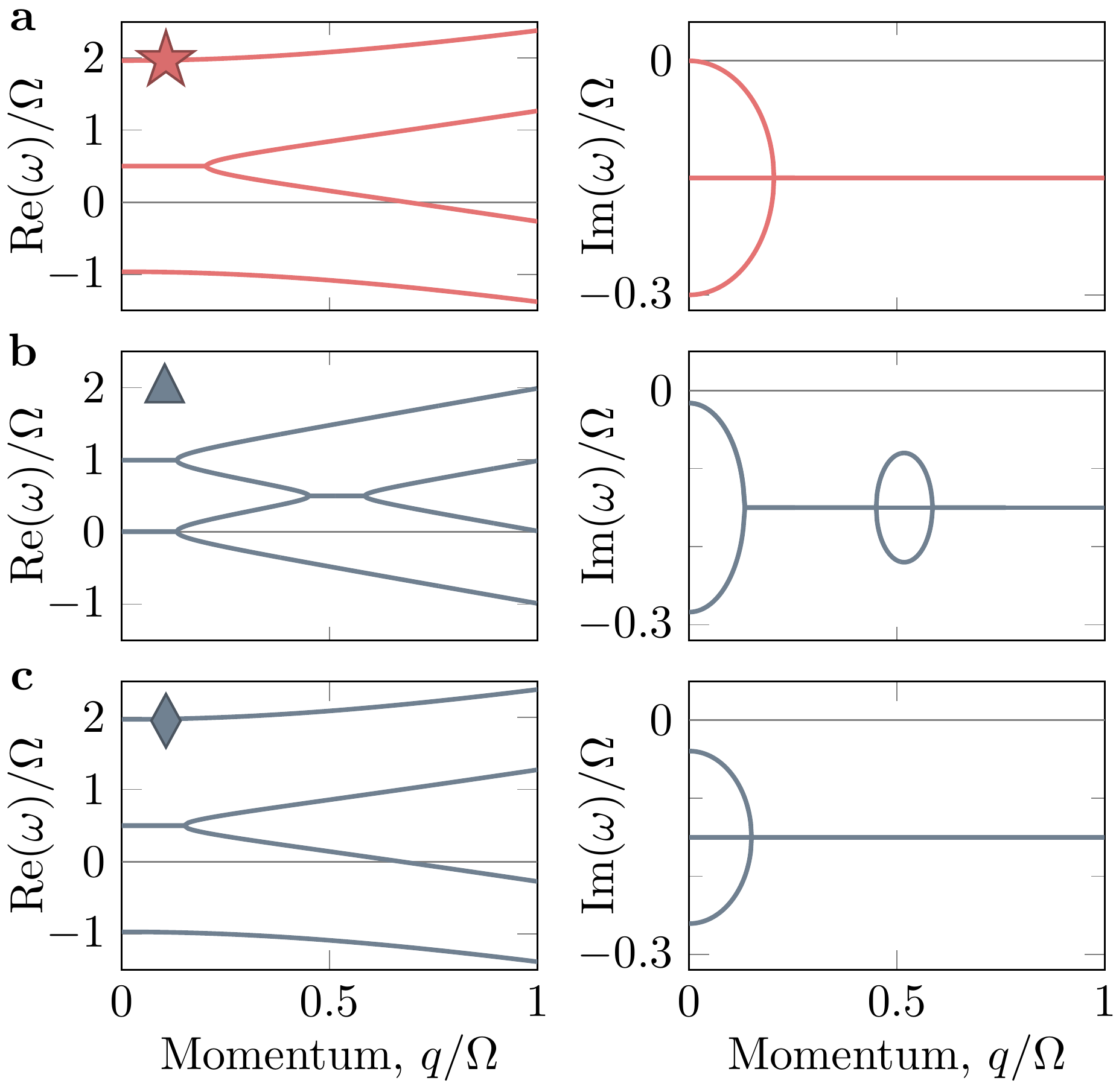}
	\caption{Real and imaginary parts of the quasiparticle dispersion relations for the two-mode model, Eq.~\eqref{eq:K kernel analyitic}. Panels (a)–(c) correspond to the parameter points marked by a star, triangle, and diamond, respectively, in the phase diagram of Fig.~\ref{Fig:OPMeanField}~(B).}
	\label{Fig:Dispersions}
\end{figure}

\subsubsection{Discussion of phase diagram and excitations}
\label{subsubsec: Excitations}
We now discuss the phase diagram emerging from the analysis, and depicted in Fig.~\ref{Fig:OPMeanField}.

\emph{Shift of critical point---} We first notice that the phase transition occurs for $r_0>0$ for finite values of $r_1$: this indicates that the presence of a drive induces superconducting phase in regions which would be normal at equilibrium.   

\emph{1st order PT and exceptional points ---} We then notice that the order parameter grows continuously from zero to a finite value across the solid black line, and discontinuously across the dashed black: This suggest that the phase transition is of the second order across the solid line, and of the first order across the dashed line. This peculiar behavior of the order parameter is typical of driven-dissipative systems which feature an exceptional point (EP) in the non-Hermitian matrix determining the order parameter~\cite{hanai2019non,hanai2020critical}.
An EP is a
point where two (or more) eigenstates that characterize the
dynamics coalesce owing to the non-Hermitian nature of the
system, such that they lose their completeness, leading to a
spectral singularity. In the present case, the matrix $\mathcal{K}(\Omega/2)$ shows an exceptional point at $r_0=\Omega^2/4$, $r_1 = \gamma\Omega/2$, i.e., where the black and the dashed lines join.  

\emph{Excitation in the normal phase---} We start from the disordered normal phase. To gain insight on the structure of the excitations, we first consider the case with $\lambda_1 = 0$. In this case the retarded Green's function describes a system of two decoupled modes (infinite decoupled modes in the case without truncation). Each mode has a complex energy with two branches (as for a damped harmonic oscillator).  
The real part of the energy is centered around either $0$ or $\Omega$ (in general around $n\Omega$ for the $n$-th mode, for the non truncated case), and they cross at $q^* = \sqrt{-r_0+(\Omega^2+\gamma^2)/4}$, provided that $r_0 < (\Omega^2+\gamma^2)/4$. The imaginary part of the energy (i.e., the inverse lifetime), instead, is the same for all modes. 
Both real and imaginary parts show a bifurcation, which is typical of driven-dissipative non-Hermitian systems~\cite{heiss2012physics}.
A finite value of the intermode coupling $\lambda_1$ creates a non-Hermitian avoided crossing of the frequencies (i.e.,  bifurcations) close to the momenta $q^*$ This emerges as a softening in the imaginary part of the energy around $q^*$ (see panel (b) bottom in Fig.~\ref{Fig:Dispersions})

By approaching the critical curves, the imaginary parts of the dispersions further softens until they become zero for some momenta, signaling the emergence of an ordered phase. The softening appears different depending on which side is approached: Upon approaching the second-order phase transition line (panel c bottom in Fig.~\ref{Fig:Dispersions}), the softening happens at $q=0$, while approaching the first-order phase transition line (panel b bottom in Fig.~\ref{Fig:Dispersions} ), the finite momentum $q^*$ softens. This is remarkable, as this would signal an instability at finite momenta, while we assumed an ordered translation-invariant phase. This suggests that our homogeneous Ansatz for the ordered phase is incomplete, and the inclusion of a spatially-modulated order parameter Ansatz would in fact show the emergence of a phase with broken translational symmetry near the first order transition line. The emergence of spatially patterned phases in driven systems is known and correspond to Faraday waves, which are observed, e.g., in periodically driven condensates, see, e.g.,~\cite{Faraday1,Faraday2,Faraday3}.
However, this goes beyond the scope of our work and we will leave it for further investigations.

\emph{Excitations in the superconducting phase---} We consider now the fluctuations in the ordered superconducting phase. In this case there is always a mode with vanishing imaginary and real part at $q=0$, which corresponds to the Goldstone mode of the ordered phase (panel a, right figure, in Fig.~\ref{Fig:Dispersions}). The mode is purely diffusive as $\text{Im}(E(q)) \propto -q^2$ for small $q$. This is well known for driven-dissipative systems and has been predicted and experimentally verified, e.g., for exciton-polariton systems~\cite{Wouters2007,claude2025observation}. Note that such mode is a manifestation of Floquet-Nambu-Goldstone modes discussed in Ref.~\cite{Sols24}.

\subsubsection{Effect of finite temperature: Fluctuations shift the critical point}

In the large frequency limit, $\Omega\gg \sqrt{|r_1|}$, $\gamma$, we can also approximate $C_n(q)\approx\frac{T}{q^2+\lambda_0}\delta_{n,0}$ and then calculate the critical value of $r_0$ for which the superconducting phase emerges.
To this end, we use the definition of $\lambda_0$ as in Eq.~\eqref{eq: lambda equation floquet} at $n=0$ and setting $b_n=0$. Then, we make use of the stability condition Eq.~\eqref{eq:stability_condition} (and replace $r_0$ with $\lambda_0$). The new critical value for $r_{0,c}$ is then given by:
\begin{multline}
	r_{0,c} = \frac{\Omega^2}{4} +\sqrt{|r_1|^2-\frac{\gamma^2\Omega^2}{4}} \\- \frac{uT}{4\pi}\log \left(1 + \frac{\Lambda^2}{\frac{\Omega^2}{4}+\sqrt{|r_1|^2-\frac{\gamma^2\Omega^2}{4}}} \right)
\end{multline}
where we performed the $\int_\qq C_0(q)$ integral and used an ultraviolet cutoff $\Lambda$ to regularize the theory. 
We therefore observe that the effect of fluctuations at finite temperatures is to reduce the critical value of $r_{0}$.

\section{Full numerical solution}
\label{sect: Full Num Solution}
In this section, we present numerical solutions of the differential equations~\eqref{eq:EoM} to show that the steady states obtained in the previous section provide qualitatively correct results.
To this end, we time evolve the system up to $t_f=9000T_0$ with $T_0=(2\pi)/\Omega$ the drive period (not to be confused with the bath temperature $T$) and analyze the order parameter $\bar{b}$ in the steady state.
The numerical phase diagram as a function of $r_{0}/\Omega^{2}$ and $r_{1}/\Omega^{2}$ is shown in Fig.~\ref{Fig:PhaseDiagram} (a) for $u=\Omega^{2}$, $\gamma=0.03\,\Omega$, $\Lambda_{\mathrm{IR}}=0.01\,\Omega$, and $\Lambda_{\mathrm{UV}}=10\,\Omega$. Three regimes appear: In the normal phase (gray), the symmetric solution $\bar{b}(t)=0$ is stable. In the superconducting (SC) phase (blue and blue hatches), an on average finite order parameter develops that is locked to the drive and is $T_0=2\pi/\Omega$–periodic, see Fig.~\ref{Fig:PhaseDiagram} (b). In the period-doubling SC region (red and red hatches), the ordered state becomes $2T_0$–periodic, breaking the discrete time-translation symmetry left by the periodic drive.
As can be seen in Fig.~\ref{Fig:PhaseDiagram} (b), in the period-doubling regions the order parameter oscillates around zero. Yet, as shown in Sec.~\ref{sec:em response}, this phase nonetheless supports a nonvanishing period-averaged Meissner effect, justifying its classification as a superconducting state.

The phase boundaries form Arnold-tongue–like lobes reminiscent of those studied for the linear Mathieu-Hill equation in Fig.~\ref{fig:Arnold tongues}. Increasing the modulation amplitude $r_{1}$ destabilizes the symmetric state and enlarges the ordered regions. The tilt and shift of the lobes with $r_{0}$ reflects the self-consistent renormalization of the effective mass $\lambda(t)$.

\textit{Finite-momentum instabilities}--- We also performed a stability analysis of each phase obtained by numerically solving Eqs.~\eqref{eq:EoM}. We found that especially for larger values of $r_1/\Omega^2$ the normal phase is unstable to a finite-momentum symmetry broken state, where the order parameter displays oscillations as a function of time with a characteristic frequency of $\Omega$ (period-preserving state) of $\Omega/2$ (period-doubling state). For details of this analysis see App.~\ref{app: finite-q}. As shown in App.~\ref{app: electromagnetic response}, such phases also exhibit a Meissner effect, which is why we dubbed them as pair density waves (PDW) in Fig.~\ref{Fig:PhaseDiagram} (marked by red or blue hatches). \change{Similarly to the analysis in Sec.~\ref{subsubsec: Excitations}, the magnitude $Q=|\bQ|$ of the momentum at which the instability occurs is approximately set by the condition $\tilde{\epsilon}_\bQ=\sqrt{Q^2+\lambda_0-\gamma^2/4}=\Omega/2$}. For a finite-momentum instability, the order parameter can take two different forms:
\begin{subequations}\label{eq: possible finite q states}
	\begin{align}
		&b_1(\mathbf{x},t)\sim \cos(\bQ\cdot\mathbf{x})f_\nu(t)\,,\quad b_{n>1}(\mathbf{x},t)=0\,,\label{eq: LO state}\\
		&[b_1+ib_2](\mathbf{x},t)\sim e^{i\bQ\cdot\mathbf{x}}f_\nu(t)\,,\quad b_{n>2}(\mathbf{x},t)=0\,,\label{eq: FF state}
	\end{align}
\end{subequations}
with $f_\nu(t)$ a time-periodic function with period $(2\pi)/\nu$. In the first case the O($N$) symmetry is broken down to O($N-1$) and both the translation and the spatial rotation symmetries are broken. In the second case, a subgroup of the combination of translations and O($N$) rotations remains preserved, while spatial rotations are broken and the global O($N$) symmetry is broken down to O($N-2$). More complex patterns are in principle possible as one is allowed to construct any linear combination of wave vectors $\bq$ satisfying $|\bq|=|\bQ|$. The determination of which finite-$q$ configuration is realized is beyond our scope and it would require solving a generalization of Eqs.~\eqref{eq:EoM} where $\bar{b}$ becomes spatially nonuniform. We also remark that our analysis (detailed in App.~\ref{app: finite-q}) relies on the assumption that the finite-$q$ order parameter develops continuously upon crossing the transition line. If the transition happens to be first-order, the boundaries of the finite-momentum phases in Fig.~\ref{Fig:PhaseDiagram} might change. We also found a regime (yellow region in Fig.~\ref{Fig:PhaseDiagram}), where $\lambda(t)$ is not a periodic function. Here, the correlation function $C_\bq(t)$ has a large peak at finite $\bq$, which is why we dubbed this phase a quasi-periodic PDW. Finite momentum instabilities in parametrically driven bosonic systems were discussed also in Refs.~\cite{Dai2021,Chandra25a,Chandra25b,Kaplan2025b}.

\section{Electromagnetic response}
\label{sec:em response}

In this section, we study the electromagnetic response of our $O(N)$-symmetric theory. 
In order to define a gauge theory, we assume $N$ to be even, and define the $N/2$ complex vector field $\phi_j=\frac{1}{\sqrt{2}}(b_j+ib_{j+N/2})$, $j=1,\dots,N/2$, and couple it to the electromagnetic vector potential $A_\mu$ via a minimal coupling
\begin{equation}\label{eq:O(N) coupled to EM field}
	\mathcal{L}=|(\partial_\mu +ie_\star A_\mu)\phi|^2-r|\phi|^2-\frac{u}{N}(|\phi|^2)^2\,, 
\end{equation}
with $e_\star$ the effective electric charge. We will be ultimately interested in $e_\star=2e$ as in the Ginzburg-Landau theory of superconductivity. Note that, due to a finite $A_\mu$, which we will treat as a perturbation in the following, in Lagrangian~\eqref{eq:O(N) coupled to EM field} the O($N$) symmetry group has been explicitly broken down to U$(N/2)\simeq$ U(1)$\times$SU($N/2$), where the U(1) part of U($N/2$) is gauged. 
We define the retarded electromagnetic kernel as
\begin{equation}
	\begin{split}
		K^{\alpha\beta}(\bq,t,t')=-i\theta(t-t')&\langle [J^\alpha_\bq(t), J^\beta_{-\bq}(t')]\rangle\\ 
		+& e_*^2 \langle |\phi(t)|^2\rangle\,\delta(t-t')\delta_{\alpha\beta}\,,
	\end{split}
\end{equation}
with $\alpha, \beta = x,y$, and 
with the current operator defined as
\begin{equation}
	J^\alpha_\bq(t)=-\frac{e_\star}{2}\int_{\boldsymbol{x}} e^{i\bq\cdot\boldsymbol{x}}\text{Im}\left[ \phi^*(\boldsymbol{x},t)\nabla_\alpha\phi(\boldsymbol{x},t)
	\right], 
\end{equation}
where $\int_{\boldsymbol{x}} \equiv\int\!d^2\boldsymbol{x}$. We then define $\mathcal{K}^{\alpha\beta}(\bq,\omega,t)$ as:
\begin{equation}
	K^{\alpha\beta}(\bq,t,t')=\int_{-\infty}^{+\infty}\!\frac{d\omega}{2\pi}\,e^{-i\omega(t-t')}\mathcal{K}^{\alpha\beta}\left(\bq,\omega,t\right)\,.
\end{equation}
In the steady state, the time dependence of $\mathcal{K}^{\alpha\beta}(\bq,\omega,t)$ will be periodic with period $T_0=2\pi/\Omega$. Moreover, rotational symmetry allows us to decompose the kernel into longitudinal and transverse components as 
\begin{align}
	\mathcal{K}^{\alpha\beta}(\bq,\omega,t)= &\frac{q_\alpha q_\beta}{q^2}\mathcal{K}_L(\bq,\omega,t) \nonumber \\ & +\left(\delta_{\alpha\beta}-\frac{q_\alpha q_\beta}{q^2}\right)\mathcal{K}_T(\bq,\omega,t).
\end{align}
\change{Expressions for the electromagnetic kernel in the steady state can be found in App.~\ref{app: electromagnetic response}.}

\subsection{Meissner effect}
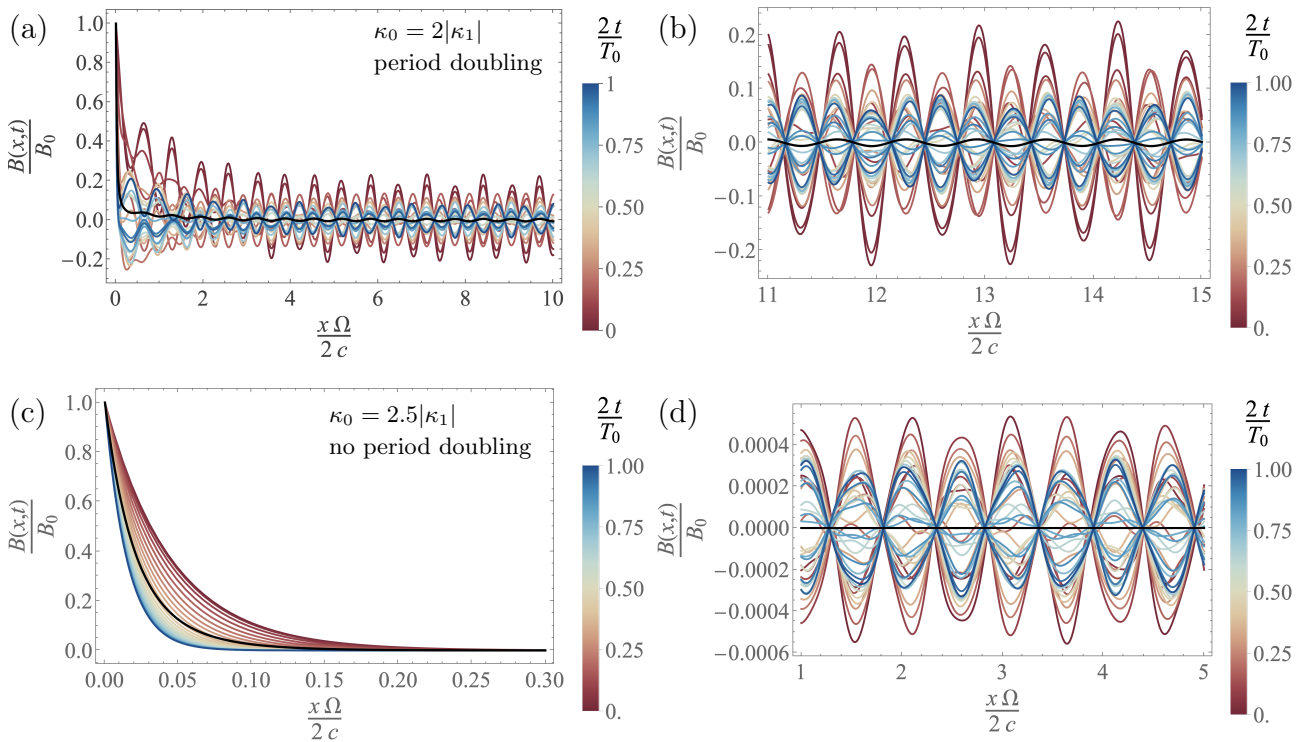
\begin{figure*}
	\centering
	\begin{overpic}[width=.475\linewidth]{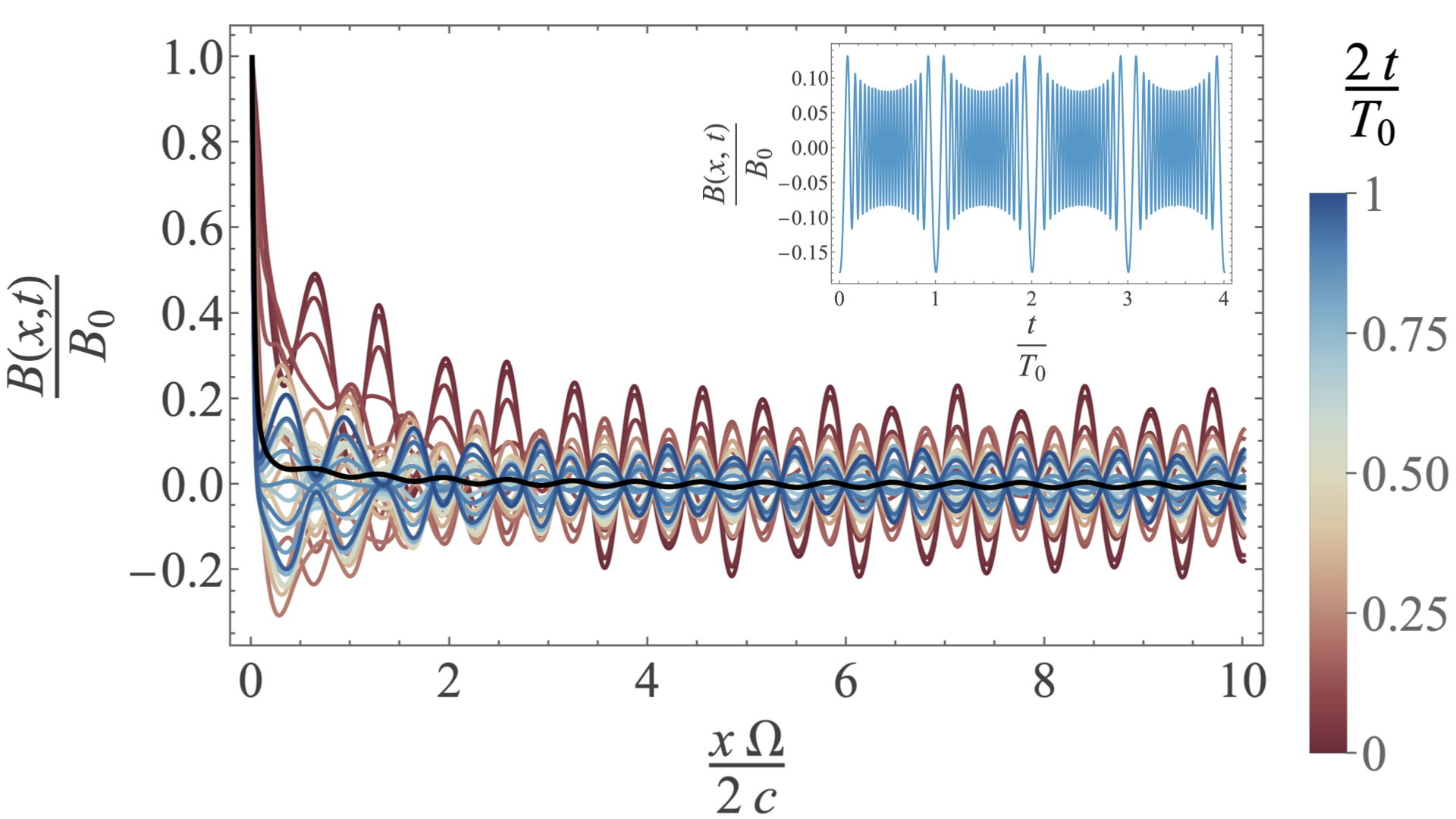}
		\put(20,50){$\kappa_0=2|\kappa_1|$}
		\put(20,45){period doubling }
		\put(0,50){\large (a)}
	\end{overpic}\vspace{2mm}
	\begin{overpic}[width=.475\linewidth]{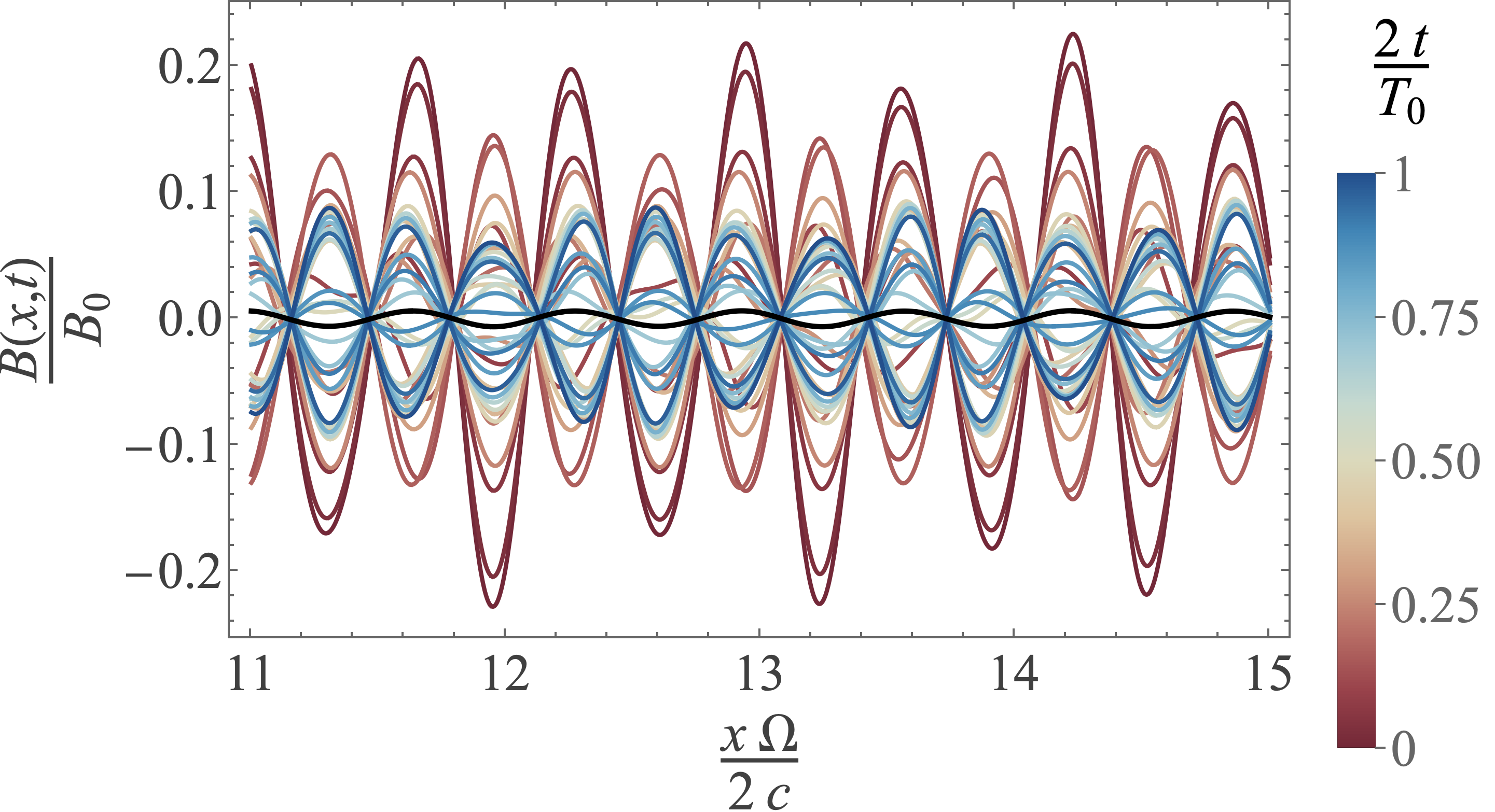}
		\put(0,50){\large (b)}
	\end{overpic}\vspace{2mm}
	\begin{overpic}[width=.475\linewidth]{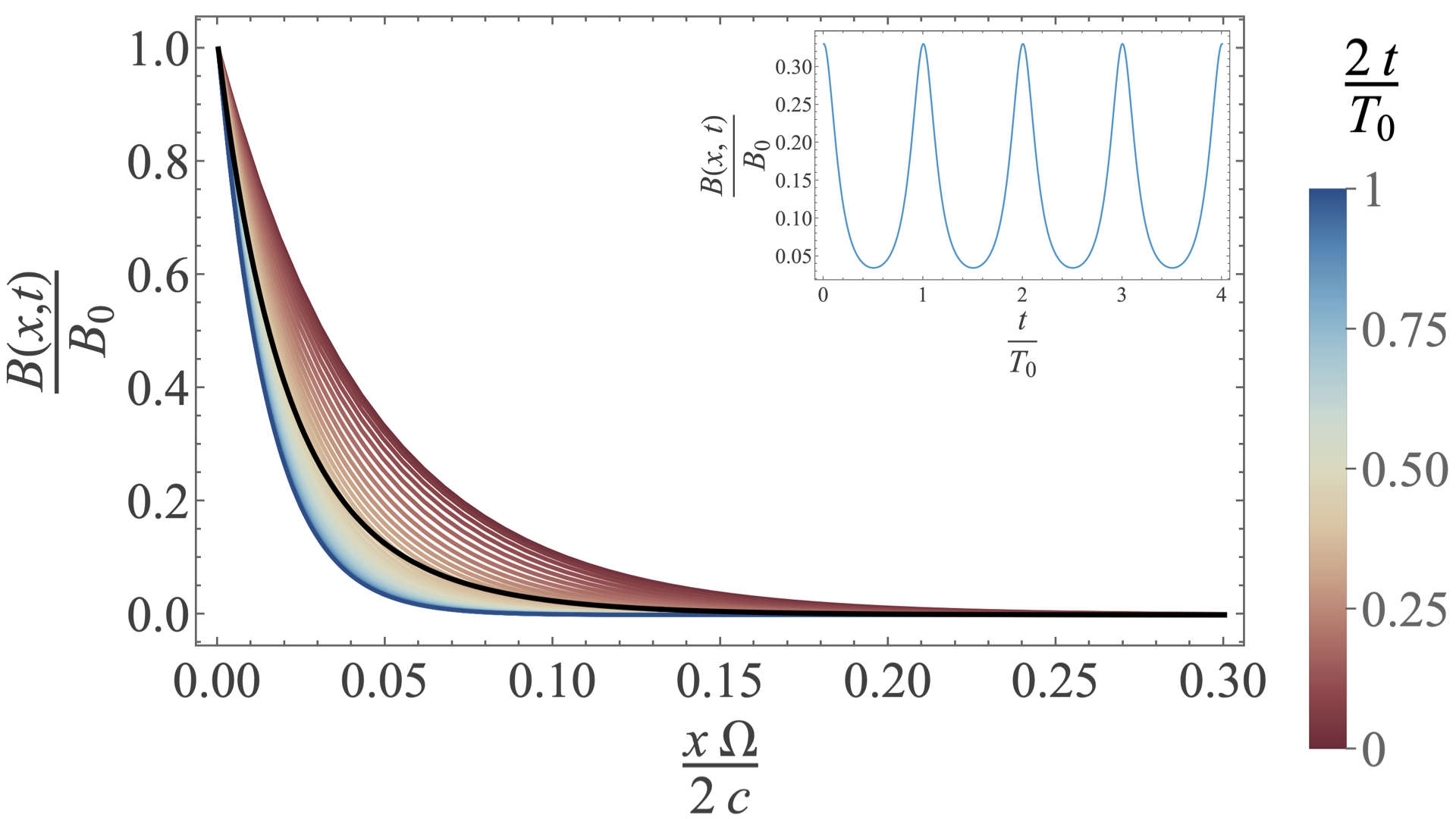}
		\put(20,50){$\kappa_0=2.5|\kappa_1|$}
		\put(20,45){no period}
		\put(33,40){doubling}
		\put(0,50){\large (c)}
	\end{overpic}\vspace{2mm}
	\begin{overpic}[width=.475\linewidth]{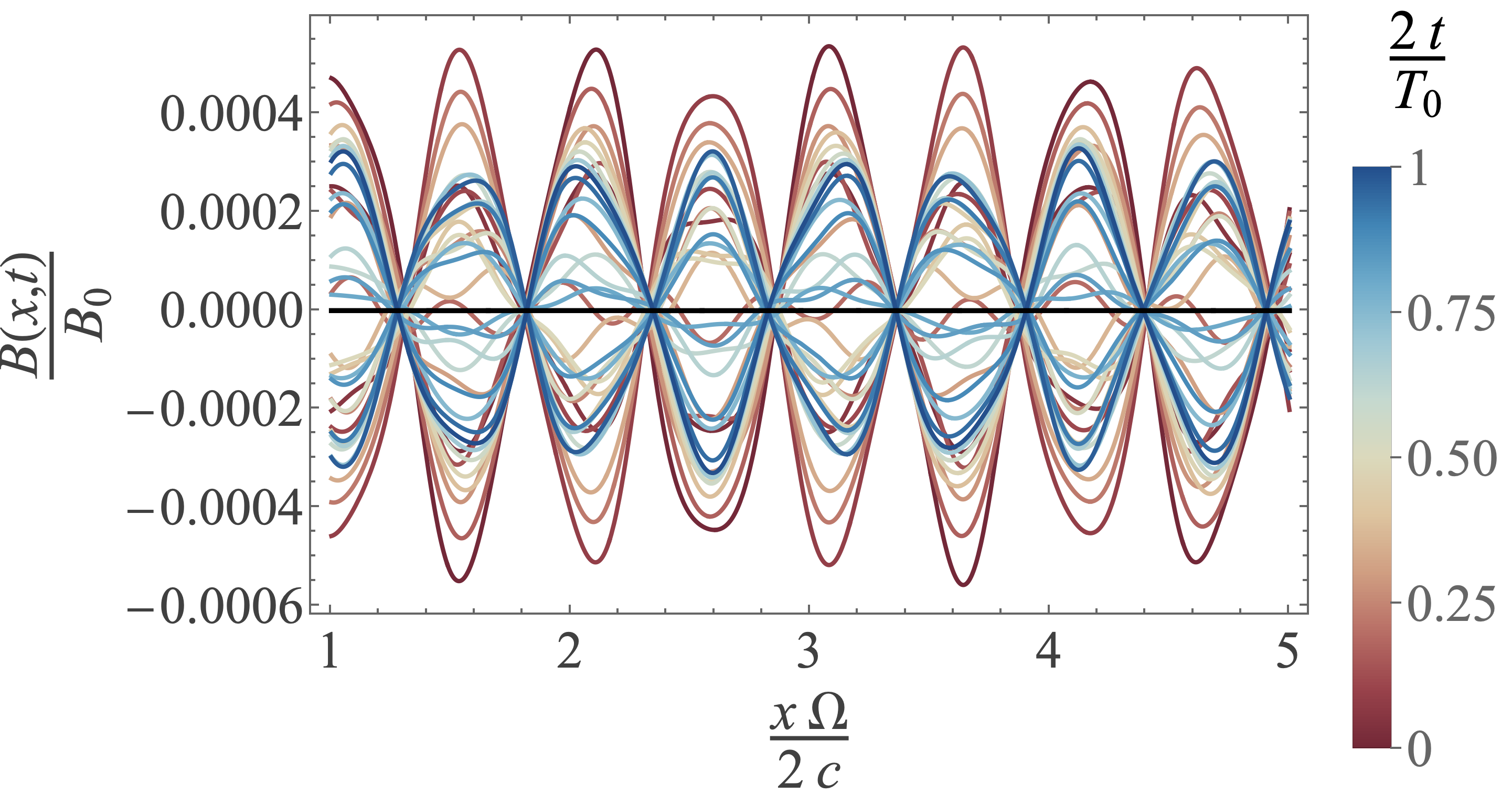}
		\put(0,50){\large (d)}
	\end{overpic}\vspace{2mm}
	\caption{Behavior of the magnetic field inside the material for a superconducting state with an order parameter that is either uniform or carries a finite momentum in the form of Eq.~\eqref{eq: FF state}. The $x$-axis coordinate indicates the distance from an interface to vacuum. In panels (a) and (b) we show a state with period doubling, whereas in panels (c) and (d) we have a superconducting state without period doubling, with an order parameter that oscillates around a finite offset. Panels (a) and (c) display the magnetic field profiles near the vacuum interface, in contrast to panels (b) and (d), which show the field behavior within the bulk of the material. \change{The insets in panels (a) and (c) show the time series of the magnetic field at $x\Omega/(2c)=10$ for $\kappa_0=2|\kappa_1|$ and at $x\Omega/(2c)=0.05$ for $\kappa_0=2.5|\kappa_1|$, respectively.} The curves have been calculated using Eq.~\eqref{eq: London time dep} for $4c^2\kappa_1/\Omega^2=1000$ for $\kappa_0=2|\kappa_1|$ (period-doubling state) and $\kappa_0=2.5|\kappa_1|$ (non period-doubling state). Different colors refer to different times in units of the half period $T_0/2=\pi/\Omega$, as indicated in the colorbars. The black curves in both panels indicate the period-averaged magnetic field.}
	\label{fig: calculated Meissner}
\end{figure*}
The Meissner effect is associated with the existence of a finite London penetration depth. Given the driven nature of the system, we should allow for the penetration length to be time-dependent. We therefore define the time-dependent London penetration depth $\lambda_L(t)$ as $\lim_{\bq\to 0}\mathcal{K}_T(\bq,\omega=0,t)\equiv c^2/(4\pi \lambda_L^2(t))$. 
As shown in App.~\ref{app: electromagnetic response}, in the large-$N$ limit, the penetration length can be directly related to the spatial average of the order parameter as  
\begin{equation}
	\label{eq:lambda_vs_orderparameter}
	\frac{c^2}{4\pi \lambda_L^2(t)}=e_\star^2 \langle|\bar{\phi}(\mathbf{x},t)|^2\rangle_\mathrm{space}\,,
\end{equation}
where $\langle\bullet\rangle_\mathrm{space}=(1/||\mathcal{V}|| )\int_{\bm x\in \mathcal{V}}\bullet(\bm x)$ denotes the spatial average over the system's volume $\mathcal{V}$, and $\bar{\phi}=\langle\phi\rangle$. 
We notice that, even in the period-doubling phases where the order parameter oscillates around zero, a penetration length with a finite time-average exist, as
\begin{equation}
	\left\langle \frac{c^2}{4\pi \lambda_L^2(t)}\right\rangle_{\!\!\!\mathrm{period}} \!\!\!\!=  e_\star^2 |\phi_0|^2\langle \cos^2(\Omega t/2)\rangle_\mathrm{period}=\frac{e_\star^2 |\phi_0|^2}{2}, 
\end{equation}
showing that the Meissner effect is expected to occur also in the period-doubling phases. Here $|\phi_0|^2=\langle |\bar{\phi}(\mathbf{x},t)|^2\rangle_\text{space,time}$.

We consider now the impact of a time-dependent London penetration depth on the shape and dynamics of the magnetic field inside the superconductor. We focus on the spatially homogeneous phases in Fig.~\ref{Fig:PhaseDiagram}, or assume the PDW state to be of the form~\eqref{eq: FF state}. In both cases we have a spatially uniform London penetration depth.


The magnetic field $\boldsymbol{B}(\boldsymbol{x},t)$ then obeys a generalized London equation of the type
\begin{equation}\label{eq: London time dep}
	\nabla^2\!\boldsymbol{B}(\boldsymbol{x},t)-\frac{1}{c^2}\frac{\partial^2 \boldsymbol{B}(\boldsymbol{x},t)}{\partial t^2}=\frac{\boldsymbol{B}(\boldsymbol{x},t)}{\lambda_L^2(t)}\,,
\end{equation}
with $c$ the speed of light in the material. 
Since $\lambda_L^2(t+2\pi/\Omega)=\lambda^2_L(t)$ for both the period-doubling and non-period-doubling phases, the previous equation can be regarded as a spatio-temporal extension of the Mathieu-Hill equation. In order to study the properties of its solutions, we work under the following assumptions. We assume that $1/\lambda_L^2(t)\approx \kappa_0-2\kappa_1 \cos(\Omega t)$, with the condition $\kappa_0\geq 2|\kappa_1|$ ensuring that $\lambda_L^2$ is always a non-negative quantity. Note that this is a simplification that we use to make analytical progress, as $\bar{b}^2(t)$ is usually a more complicated function. We expect, however, our results not to be qualitatively affected by such a choice.
We then assume the sample to be a semi-infinite slab with a planar interface between vacuum for $x<0$ and a driven superconductor for $x>0$, with a uniform and static magnetic field $\boldsymbol{B}_0$ in the vacuum region.
In this case, the problem becomes effectively one-dimensional, with boundary conditions $\boldsymbol{B}(x=0,t)=\boldsymbol{B}_0$, $\partial\boldsymbol{B}(x,t)/\partial t|_{t=0}=0$. Furthermore, we require the magnetic field to be bounded $|\mathrm{sup}_{x,t}\boldsymbol{B}(x,t)|<\infty$.

Within these assumptions, Eq.~\eqref{eq: London time dep} admits the following close solution
\begin{equation}
	\label{eq: B(x,t)}
	\boldsymbol{B}(x,t)=\boldsymbol{B}_0\sum_{n=0}^{+\infty} f_n\,\Phi_n\!\left(x\right)\,\mathrm{ce}_{2n}\!\!\left(\frac{4c^2\kappa_1}{\Omega^2},\frac{\Omega t}{2}\right),
\end{equation}
with $\mathrm{ce}_n(y,\tau)$ time-periodic cosine-Mathieu functions of the first kind~\cite{MathieuNote}, and the $\Phi_n(x)$ functions defined as 
\begin{equation}
	\label{eq:phi_space_dependent}
	\Phi_n(x)=
	\begin{cases}
		e^{-\sqrt{\alpha_n}x} & \text{if } \alpha_n > 0,\\[4pt]
		1 & \text{if } \alpha_n = 0,\\[4pt]
		\frac{\cos(\sqrt{-\alpha_n}\,x+\theta_n)}{\cos\theta_n} & \text{if } \alpha_n < 0,
	\end{cases}
\end{equation}
with $\theta_n\in(-\pi/2,\pi/2)$ undetermined, as the problem with the present boundary conditions is underdetermined, 
and $\alpha_n$ defined as $\alpha_n= \kappa_0-[\Omega/(2c)]^2\,a_{2n}(4c^2 \kappa_1/\Omega^2)$, with $a_n(y)$ the type-A Mathieu characteristic numbers. Finally, the coefficients $f_n$ are calculated as 
\begin{equation}
	\label{eq:fn_coefficient}
	f_n=\frac{2}{\pi}\int_0^\pi\!d\tau\,\mathrm{ce}_{2n}\!\!\left(\frac{4c^2\kappa_1}{\Omega^2},\tau\right).
\end{equation}
\begin{figure*}[t]
	\centering
	\includegraphics[width=1\linewidth]{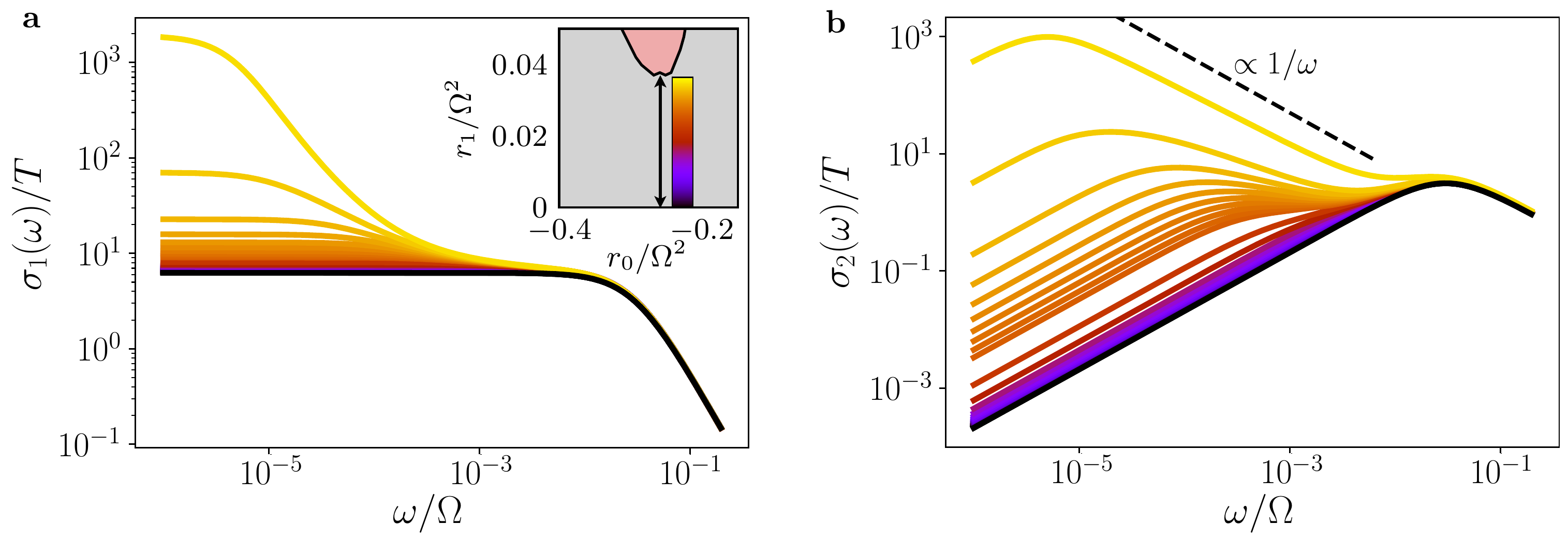}
	\caption{Real (panel (a)) and imaginary (panel (b)) part of the optical conductivity  in units of the bath temperature $T$ for $r_0=-0.25\,\Omega^2$ for different values of $r_1/\Omega^2$ approaching the bottom of the uniform superconducting phase with period doubling in Fig.~\ref{Fig:PhaseDiagram}, as shown in the inset of panel (a). All curves have been calculated in the normal phase. In panel (b) the black dashed line represents the $1/\omega$ behavior expected in a superconductor.}
	\label{fig: optical conductivity}
\end{figure*}
The magnetic field inside the driven superconductor oscillates periodically with the same period of the drive. The main difference compared to the equilibrium case comes from the contributions of the spatial dependence of the functions~\eqref{eq:phi_space_dependent}: depending on the value of $\alpha_n$, which is controlled by the form of the time-dependent London penetration length $\lambda_L(t)$, different spatial profiles are possible, i.e., exponentially decaying (as in the equilibrium case), or indefinitely oscillating. 
This implies that, in a driven superconductor, magnetic fields with finite spatial and temporal modulation can penetrate, in stark contrast to the strict expulsion expected in equilibrium.

The natural question is then how large these oscillations actually are. 
Their magnitude is set by the interplay between the eigenvalues $\alpha_n$ 
in Eq.~\eqref{eq:phi_space_dependent} and the coefficients $f_n$ in 
Eq.~\eqref{eq:fn_coefficient}. For small $n$, the $\alpha_n$ are positive, 
whereas for larger $n$ they become increasingly negative. In contrast, the 
coefficients $f_n$ decrease rapidly with $n$. As a result, the low-$n$ modes 
always produce spatially decaying profiles, while spatially oscillatory 
profiles appear only at sufficiently large $n$; however, their weight is 
strongly suppressed by the small values of $f_n$. 
We numerically observed that for a state with period doubling ($\kappa_0=2|\kappa_1|$) there is typically only one value of $n=\bar{n}$ so that $\alpha_{\bar{n}}<0$ and $f_{\bar{n}}$ is not yet exponentially suppressed. In this case, the magnitude of the oscillations is given by $|f_{\bar{n}}|\,\mathrm{max}_{\tau}|\mathrm{ce}_{2\bar{n}}(4c^2\kappa_1/\Omega^2,\tau)|$ and their wavelength is given by $(2\pi)/\sqrt{-\alpha_{\bar{n}}}$. The time-averaged amplitude of the spatial oscillations is instead given by $f^2_{\bar{n}}/2$. For the superconducting phase without period doubling that is continuously connected to the equilibrium superconducting phase (bottom left of Fig.~\ref{Fig:PhaseDiagram}), we have $\kappa_0\gg 2|\kappa_1|$, which, compared to the case with period doubling, gives a large positive shift to all $\alpha_n$, making it impossible to find an $n=\bar{n}$ where $\alpha_{\bar{n}<0}$ and $f_{\bar{n}}$ is not exponentially suppressed. We therefore recover the conventional Meissner effect in this case, with a complete expulsion of the magnetic field from the material. In the other uniform superconducting phase without period doubling in Fig.~\ref{Fig:PhaseDiagram}, we expect $\bar{b}(t)\approx \cos(\Omega t)$, implying $\kappa_0\gtrsim 2|\kappa_1|$, making it qualitatively similar to the period-doubling phase (with the replacement $\Omega\to 2\Omega)$.

In Fig.~\ref{fig: calculated Meissner}, we show the space and time dependence of $\boldsymbol{B}(\boldsymbol{x},t)$ in a period-doubling phase $\kappa_0=2|\kappa_1|$ (panels (a-b)). 
We choose parameters consistent with current experimental conditions. 
Specifically, we take $\kappa_1 \approx 1/(100~\mathrm{nm})^2$, with 
$100~\mathrm{nm}$ a representative value of the London penetration depth, 
and we set $\Omega \approx 100~\mathrm{THz}$, a frequency scale typical of 
ultrafast pump--probe experiments. Assuming that the speed of light is not significantly renormalized in the material, these values give 
$4c^2 \kappa_1 / \Omega^2 \approx 10^3$–$10^4$. 
In the following, we consider the representative choice 
$4c^2 \kappa_1 / \Omega^2 = 10^3$.
We observe that the applied magnetic field penetrates the oscillating superconductor in the form of a standing wave, with an amplitude that corresponds to about 20\% of the externally applied field (panel(b)).
As the experimental setups may be unable to resolve fast oscillations, in Fig.~\ref{fig: calculated Meissner}, we also show the magnetic field averaged over a period of the external drive (black curves). Even on average, the magnetic field penetrates the sample and forms a spatially oscillating pattern with a wavelength that depends on $\kappa_1$. The fraction of static magnetic field that penetrates sample is, however quite small, 0.6\% of the applied field for the present choice of parameters. In panels (c-d) of Fig.~\ref{fig: calculated Meissner}, we show the behavior of the magnetic field for a state without period doubling, that is allowed to have an order parameter oscillating around a finite offset. We choose here $\kappa_0=2.5|\kappa_1|$, corresponding to a state where the order parameter never vanishes. We observe that close to the interface the magnetic field displays some oscillations but it eventually decays to zero at larger distances. On average the spatial profile resembles that of an ordinary Meissner effect in this case. Deep in the material oscillations of the magnetic field persist also in this case (panel (d) of Fig.~\ref{fig: calculated Meissner}), but with a strongly suppressed amplitude, that further decreases by increasing $\kappa_0$. Such oscillations, however, average to zero (black curve in panel (d) of Fig.~\ref{fig: calculated Meissner}).

Finally, we note that in both the same-period and period-doubled phases (Fig.~\ref{fig: calculated Meissner}, insets of panels (a) and (c)), the magnetic field $\boldsymbol{B}(\boldsymbol{x},t)$ oscillates with period $T_0$. This follows from the fact that the inverse squared penetration depth is proportional to $|\phi(t)|^2$, see Eq.~\eqref{eq:lambda_vs_orderparameter}. While $\phi(t)$ can be $T_0$- or $2T_0$-periodic depending on whether $\nu=0$ or $\nu=1/2$ in Eq.~\eqref{eq: Floquet b}, its square $|\phi(t)|^2$ is always $T_0$-periodic. In fact, if it is $T_0$ periodic, it will always contain a time-independent component (see Eq.~\eqref{eq: Floquet b}), whereas this vanishes in period-doubling phases. Consequently, $\boldsymbol{B}(\boldsymbol{x},t)$ inherits the fundamental period $T_0$ in both phases. We also note that, in the theory in Eq.~\eqref{eq:O(N) coupled to EM field}, all gauge invariant quantities must contain $\phi(t)$ as $|\phi(t)|^2$, making it impossible to observe $2T_0$ oscillations. Conversely, if the theory in Eq.~\eqref{eq: b potential} is not coupled to a gauge field and if the order parameter $b(t)$ itself is an observable, $2T_0$ oscillations are in principle possible to detect. 

We have therefore shown that if the order parameter oscillates around a sufficiently small (for non-period doubling phases) or vanishing (period-doubling phases) offset, the magnetic field can penetrate the sample, realizing a standing wave whose amplitude is a fraction of the externally applied field. Spatial modulations of the field survive also when a time average is performed, although with a further suppressed amplitude. These oscillations can be thought of as resulting from a collective mode emerging from the coupling of light to order parameter oscillations, which we dub as \textit{Meissner polariton}.

\subsection{Optical conductivity}
The period-averaged optical conductivity is defined in terms of the electromagnetic response kernel ${K}^{\alpha\alpha}$ as  
\begin{equation}
	\sigma(\omega)=\frac{\langle\mathcal{K}^{\alpha\alpha}(\bq=0,\omega,t)\rangle_\mathrm{period}}{i\omega}\,.
\end{equation}
The computation of $\sigma(\omega)$ is detailed in App.~\ref{app: electromagnetic response}, and we report here only report the numerical results. In Fig.~\ref{fig: optical conductivity}, we show the real and imaginary part of the period-averaged optical conductivity for values of $r_1$ in the range $[0,\approx\!0.04\,\Omega^2]$, all corresponding to a normal phase in Fig.~\ref{Fig:PhaseDiagram} (see also inset here for reference on the color scale).

We first note that, although the system is in the normal gapped phase, 
the combination of the high-temperature approximation and a finite 
dissipation rate $\gamma$ leads to the appearance of a Drude peak already 
in equilibrium (black curve).
Upon increasing the drive amplitude $r_1$, the system approaches a period-doubling uniform superconducting phase, and the optical conductivity starts to strongly deviate from being Drude-like. The real part starts developing a high narrow peak around $\omega=0$, while the imaginary part displays a $1/\omega$ behavior over a range of frequencies that increases as $r_1$ grows.
Upon entering the period-doubling uniform superconducting phase (not shown in here), the optical conductivity picks up an additional contribution which takes the form of a $\delta$-function for the real part and of a $1/\omega$ decay in the imaginary one, similarly to what happens in a superconductor at equilibrium. The prefactor of this additional contribution is proportional to $\bar{b}^2(t)$ averaged over a period of the drive. 

In summary, Fig.~\ref{fig: optical conductivity} shows that it is possible to achieve a superconducting-like response (that is, $\sigma_2\sim1/\omega$) for several orders of magnitude in $\omega$ even in the absence of a Meissner effect, provided that one is proximate to a non-equilibrium phase with an oscillating order parameter.

\section{Conclusion}
\label{sec: conclusions}
In this work, we have mapped out the phase diagram of a parametrically driven relativistic O($N$)-symmetric model coupled to a thermal bath, using Floquet and Keldysh field theory and combining analytical and numerical approaches. Our analysis reveals a rich phase diagram, including states with spontaneously broken O($N$) symmetry emerging from a symmetric, disordered phase upon increasing the drive strength. Beyond the global O($N$) symmetry, these phases can also exhibit the breaking of additional symmetries, such as discrete time-translation or spatial rotation and translation symmetry.

For even values of $N$, we coupled the system to a background electromagnetic vector potential and demonstrated that all equilibrium and non-equilibrium phases with broken O($N$) symmetry exhibit a Meissner effect, in the sense that the electromagnetic field acquires a mass. Remarkably, when discrete time-translation symmetry is broken and an external magnetic field is applied, a portion of the magnetic flux can penetrate the system, forming a standing-wave pattern within the sample. We dubbed such pattern a \textit{Meissner polariton}, a collective mode emerging from the interaction of order parameter oscillations with light. \change{This effect can be detected with a magnetometer with sufficiently high spatial resolution. In fact, as shown in Fig.~\ref{fig: calculated Meissner}(b), the magnetic field penetrates the sample in the form of a standing wave, and a spatially oscillating profile survives the time-average over the fast oscillations. We estimate the wave length of the spatial, period-averaged, profile to be of the order of $\lambda \approx c/\Omega$, which for $\Omega\approx 100$ THz gives $\lambda\approx 3$ $\mu$m. A magnetometer with micrometric (or higher) resolution would therefore be needed to detect such effect. }
Furthermore, we found that increasing the drive strength can strongly enhance the imaginary part of the optical conductivity, even in the absence of O($N$) symmetry breaking or a Meissner effect.

Our results have broad implications. They establish that dynamical driving can induce emergent ordered phases, offering a unified theoretical framework for light-induced symmetry breaking phenomena observed not only in superconductors, but also in charge-density-wave, antiferromagnetic, and ferroelectric systems. Moreover, we showed that drive-induced symmetry-broken phases can display a time-dependent Meissner effect, providing a potential explanation for the recently reported magnetic field expulsion in optically driven YBCO~\cite{Cavalleri24}. 
We also demonstrated that near a non-equilibrium superconducting phase, the imaginary part of the optical conductivity can exhibit a pronounced $1/\omega$ scaling over several frequency decades, mimicking superconducting behavior even without a genuine Meissner response.

Promising directions for future work include applying our framework to microscopic models with explicit electronic degrees of freedom, and extending it to more complex field theories that feature the competition or coexistence of multiple order parameters~\cite{Diessel_inprep}.

\subsection*{Acknowledgments}
We gratefully acknowledge discussions A.~Cavalleri, A.~Chiocchetta, E.~Demler, J.S.~Dodge, P.E.~Dolgirev, F.~Glerean, M.~Mitrano, P.A.~Volkov, and C.P.~Zelle. This research was supported by NSF Grant DMR-2245246 and by the Simons Collaboration on Ultra-Quantum Matter which is a grant from the Simons Foundation (651440, S. S.). 
The Flatiron Institute is a division of the Simons Foundation.
O.K.D acknowledges support from the NSF through a grant for ITAMP at Harvard University.
P.M.B. acknowledges support by the German National Academy of Sciences Leopoldina through Grant No.~LPDS 2023-06 and the Gordon and Betty Moore Foundation’s EPiQS Initiative Grant GBMF8683.
\bibliography{biblio}

@ARTICLE{Comin25,
	author = {{Kang}, Mingu and {Dolgirev}, Pavel E. and {Zhang}, Chao C. and {Jang}, Hoyoung and {Lee}, Byungjune and {Kim}, Minseok and {Park}, Sang-Youn and {Sutarto}, Ronny and {Demler}, Eugene and {Park}, Jae-Hoon and {Wei}, John Y.~T. and {Comin}, Riccardo},
	title = "{Role of Phase Fluctuation in Dynamic Competition Between Charge Order and Superconductivity in Cuprates}",
	journal = {arXiv e-prints},
	year = 2025,
	month = oct,
	eid = {arXiv:2510.26763},
	pages = {arXiv:2510.26763},
	doi = {10.48550/arXiv.2510.26763},
	archivePrefix = {arXiv},
	eprint = {2510.26763},
	primaryClass = {cond-mat.str-el},
	adsurl = {https://ui.adsabs.harvard.edu/abs/2025arXiv251026763K}
}

@ARTICLE{Zelle25,
	author = {{Daviet}, Romain and {Zelle}, Carl Philipp and {Asadollahi}, Armin and {Diehl}, Sebastian},
	title = "{Kardar-Parisi-Zhang Scaling in Time-Crystalline Matter}",
	journal = {Phys. Rev. Lett.},
	year = 2025,
	month = jul,
	volume = {135},
	number = {4},
	eid = {047101},
	pages = {047101},
	doi = {10.1103/pbtn-wsgv},
	archivePrefix = {arXiv},
	eprint = {2412.09677},
	primaryClass = {cond-mat.stat-mech},
	adsurl = {https://ui.adsabs.harvard.edu/abs/2025PhRvL.135d7101D}
}

@ARTICLE{Gedik25a,
	author = {{Su}, Yifan and {Lv}, B.~Q. and {Zong}, Alfred and {M\"uller}, Aaron and {Chattopadhyay}, Sambuddha and {Dolgirev}, Pavel E. and {Singh}, Anisha G. and {Straquadine}, Joshua A.~W. and {Choi}, Dongsung and {Azoury}, Doron and {Mogi}, Masataka and {Fisher}, Ian R. and {Demler}, Eugene and {Gedik}, Nuh},
	title = "{Time-domain identification of distinct mechanisms for competing charge density waves in a rare-earth tritelluride}",
	journal = {arXiv e-prints},
	year = 2025,
	month = mar,
	eid = {arXiv:2503.13936},
	pages = {arXiv:2503.13936},
	doi = {10.48550/arXiv.2503.13936},
	archivePrefix = {arXiv},
	eprint = {2503.13936},
	primaryClass = {cond-mat.str-el},
	adsurl = {https://ui.adsabs.harvard.edu/abs/2025arXiv250313936S}
}

@ARTICLE{Gedik25b,
	author = {{Ning}, Honglie and {Oh}, Kyoung Hun and {Su}, Yifan and {Darius Shi}, Zhengyan and {Wu}, Dong and {Liu}, Qiaomei and {Lv}, B.~Q. and {Zong}, Alfred and {Kang}, Gyeongbo and {Choi}, Hyeongi and {Kim}, Hyun-Woo J. and {Ha}, Seunghyeok and {Kim}, Jaehwon and {Sarker}, Suchismita and {Ruff}, Jacob P.~C. and {Kim}, B.~J. and {Wang}, N.~L. and {Senthil}, Todadri and {Jang}, Hoyoung and {Gedik}, Nuh},
	title = "{Bidirectional ultrafast control of charge density waves via phase competition}",
	journal = {arXiv e-prints},
	year = 2025,
	month = sep,
	eid = {arXiv:2510.00204},
	pages = {arXiv:2510.00204},
	doi = {10.48550/arXiv.2510.00204},
	archivePrefix = {arXiv},
	eprint = {2510.00204},
	primaryClass = {cond-mat.str-el},
	adsurl = {https://ui.adsabs.harvard.edu/abs/2025arXiv251000204N}
}

@ARTICLE{Gedik24,
	author = {{Ilyas}, Batyr and {Luo}, Tianchuang and {von Hoegen}, Alexander and {Vi{\~n}as Bostr{\"o}m}, Emil and {Zhang}, Zhuquan and {Park}, Jaena and {Kim}, Junghyun and {Park}, Je-Geun and {Nelson}, Keith A. and {Rubio}, Angel and {Gedik}, Nuh},
	title = "{Terahertz field-induced metastable magnetization near criticality in FePS$_{3}$}",
	journal = {\nat},
	year = 2024,
	month = dec,
	volume = {636},
	number = {8043},
	pages = {609-614},
	doi = {10.1038/s41586-024-08226-x},
	archivePrefix = {arXiv},
	eprint = {2507.06371},
	primaryClass = {cond-mat.mtrl-sci},
	adsurl = {https://ui.adsabs.harvard.edu/abs/2024Natur.636..609I}
}

@ARTICLE{Dodge25,
	author = {{Dodge}, J. Steven and {Lopez}, Leya and {Sahota}, Derek G.},
	title = "{Status of the Spurious Evidence for Photoinduced Superconductivity}",
	journal = {arXiv e-prints},
	year = 2023,
	month = jun,
	eid = {arXiv:2307.00204},
	pages = {arXiv:2307.00204},
	doi = {10.48550/arXiv.2307.00204},
	archivePrefix = {arXiv},
	eprint = {2307.00204},
	primaryClass = {cond-mat.supr-con},
	adsurl = {https://ui.adsabs.harvard.edu/abs/2023arXiv230700204D}
}

@ARTICLE{Xu25,
	author = {{Xu}, Shuxiang and {Wang}, Hao and {Huo}, Mengwu and {Hu}, Deyuan and {Wu}, Qiong and {Yue}, Li and {Wu}, Dong and {Wang}, Meng and {Dong}, Tao and {Wang}, Nanlin},
	title = "{Collapse of density wave and emergence of superconductivity in pressurized-La$_{4}$Ni$_{3}$O$_{10}$ evidenced by ultrafast spectroscopy}",
	journal = {Nature Communications},
	year = 2025,
	month = jul,
	volume = {16},
	number = {1},
	eid = {7039},
	pages = {7039},
	doi = {10.1038/s41467-025-62294-9},
	archivePrefix = {arXiv},
	eprint = {2503.05176},
	primaryClass = {cond-mat.supr-con},
	adsurl = {https://ui.adsabs.harvard.edu/abs/2025NatCo..16.7039X}
}

@ARTICLE{Diehl16,
	author = {{Sieberer}, L.~M. and {Buchhold}, M. and {Diehl}, S.},
	title = "{Keldysh field theory for driven open quantum systems}",
	journal = {Reports on Progress in Physics},
	year = 2016,
	month = sep,
	volume = {79},
	number = {9},
	eid = {096001},
	pages = {096001},
	doi = {10.1088/0034-4885/79/9/096001},
	archivePrefix = {arXiv},
	eprint = {1512.00637},
	primaryClass = {cond-mat.quant-gas},
	adsurl = {https://ui.adsabs.harvard.edu/abs/2016RPPh...79i6001S}
}

@ARTICLE{ZelleMillis25,
	author = {{Zelle}, Carl Philipp and {Daviet}, Romain and {Millis}, Andrew J. and {Diehl}, Sebastian},
	title = "{Nonequilibrium orders in parametrically driven field theories}",
	journal = {arXiv e-prints},
	year = 2025,
	month = jun,
	eid = {arXiv:2506.18622},
	pages = {arXiv:2506.18622},
	doi = {10.48550/arXiv.2506.18622},
	archivePrefix = {arXiv},
	eprint = {2506.18622},
	primaryClass = {cond-mat.stat-mech},
	adsurl = {https://ui.adsabs.harvard.edu/abs/2025arXiv250618622Z}
}

@ARTICLE{Dai22,
	author = {{Dai}, Zhehao and {Ravindran}, Vibhu and {Yao}, Norman Y. and {Zaletel}, Michael P.},
	title = "{Photo-induced Superconductivity = Discrete Time Crystal?}",
	journal = {arXiv e-prints},
	year = 2022,
	month = sep,
	eid = {arXiv:2209.05510},
	pages = {arXiv:2209.05510},
	doi = {10.48550/arXiv.2209.05510},
	archivePrefix = {arXiv},
	eprint = {2209.05510},
	primaryClass = {cond-mat.supr-con},
	adsurl = {https://ui.adsabs.harvard.edu/abs/2022arXiv220905510D}
}

@ARTICLE{Cremin19,
	author = {{Cremin}, Kevin A. and {Zhang}, Jingdi and {Homes}, Christopher C. and {Gu}, G.~D. and {Sun}, Zhiyuan and {Fogler}, Michael M. and {Millis}, Andrew J. and {Basov}, D.~N. and {Averitt}, Richard D.},
	title = "{Photoenhanced metastable c-axis electrodynamics in stripe-ordered cuprate La$_{1.885}$Ba$_{0.115}$CuO$_{4}$}",
	journal = {Proceedings of the National Academy of Science},
	year = 2019,
	month = oct,
	volume = {116},
	number = {40},
	pages = {19875-19879},
	doi = {10.1073/pnas.1908368116},
	archivePrefix = {arXiv},
	eprint = {1901.10037},
	primaryClass = {cond-mat.supr-con},
	adsurl = {https://ui.adsabs.harvard.edu/abs/2019PNAS..11619875C}
}

@ARTICLE{DaiLee21,
	author = {{Dai}, Zhehao and {Lee}, Patrick A.},
	title = "{Superconducting-like response in driven systems near the Mott transition}",
	journal = {Phys. Rev. B},
	year = 2021,
	month = dec,
	volume = {104},
	number = {24},
	eid = {L241112},
	pages = {L241112},
	doi = {10.1103/PhysRevB.104.L241112},
	archivePrefix = {arXiv},
	eprint = {2106.08354},
	primaryClass = {cond-mat.str-el},
	adsurl = {https://ui.adsabs.harvard.edu/abs/2021PhRvB.104x1112D}
}

@ARTICLE{Comment23,
	author = {{Buzzi}, M. and {Nicoletti}, D. and {Rowe}, E. and {Wang}, E. and {Cavalleri}, A.},
	title = "{Comment on arXiv:2210.01114: Optical Saturation Produces Spurious Evidence for Photoinduced Superconductivity in K$_3$C$_{60}$}",
	journal = {arXiv e-prints},
	year = 2023,
	month = mar,
	eid = {arXiv:2303.10169},
	pages = {arXiv:2303.10169},
	doi = {10.48550/arXiv.2303.10169},
	archivePrefix = {arXiv},
	eprint = {2303.10169},
	primaryClass = {cond-mat.supr-con},
	adsurl = {https://ui.adsabs.harvard.edu/abs/2023arXiv230310169B}
}

@ARTICLE{Chandra25a,
	author = {{Kaplan}, Daniel and {Volkov}, Pavel A. and {Chakraborty}, Ahana and {Zhuang}, Zekun and {Chandra}, Premala},
	title = "{Tunable Spatiotemporal Orders in Driven Insulators}",
	journal = {Phys. Rev. Lett.},
	year = 2025,
	month = feb,
	volume = {134},
	number = {6},
	eid = {066902},
	pages = {066902},
	doi = {10.1103/PhysRevLett.134.066902},
	archivePrefix = {arXiv},
	eprint = {2405.12214},
	primaryClass = {cond-mat.mes-hall},
	adsurl = {https://ui.adsabs.harvard.edu/abs/2025PhRvL.134f6902K}
}

@ARTICLE{Chandra25b,
	author = {{Kaplan}, Daniel and {Volkov}, Pavel A. and {Coulter}, Jennifer and {Zhang}, Shiwei and {Chandra}, Premala},
	title = "{Spatiotemporal Order and Parametric Instabilities from First-Principles}",
	journal = {arXiv e-prints},
	year = 2025,
	month = jul,
	eid = {arXiv:2507.14110},
	pages = {arXiv:2507.14110},
	doi = {10.48550/arXiv.2507.14110},
	archivePrefix = {arXiv},
	eprint = {2507.14110},
	primaryClass = {cond-mat.mtrl-sci},
	adsurl = {https://ui.adsabs.harvard.edu/abs/2025arXiv250714110K}
}

@ARTICLE{Dodge23,
	author = {{Dodge}, J. Steven and {Lopez}, Leya and {Sahota}, Derek G.},
	title = "{Optical Saturation Produces Spurious Evidence for Photoinduced Superconductivity in K$_{3}$C$_{60}$}",
	journal = {Phys. Rev. Lett.},
	year = 2023,
	month = apr,
	volume = {130},
	number = {14},
	eid = {146002},
	pages = {146002},
	doi = {10.1103/PhysRevLett.130.146002},
	archivePrefix = {arXiv},
	eprint = {2210.01114},
	primaryClass = {cond-mat.supr-con},
	adsurl = {https://ui.adsabs.harvard.edu/abs/2023PhRvL.130n6002D}
}

@ARTICLE{Sols24,
	author = {{Mu{\~n}oz de Nova}, Juan Ram{\'o}n and {Sols}, Fernando},
	title = "{Simultaneous symmetry breaking in spontaneous Floquet states: temporal Floquet-Nambu-Goldstone modes, Floquet thermodynamics, and the time operator}",
	journal = {Quantum},
	year = 2025,
	month = sep,
	volume = {9},
	pages = {1850},
	doi = {10.22331/q-2025-09-05-1850},
	archivePrefix = {arXiv},
	eprint = {2402.10784},
	primaryClass = {quant-ph},
	adsurl = {https://ui.adsabs.harvard.edu/abs/2025Quant...9.1850M}
}

@ARTICLE{Gedik21,
	author = {{Zong}, Alfred and {Dolgirev}, Pavel E. and {Kogar}, Anshul and {Su}, Yifan and {Shen}, Xiaozhe and {Straquadine}, Joshua A.~W. and {Wang}, Xirui and {Luo}, Duan and {Kozina}, Michael E. and {Reid}, Alexander H. and {Li}, Renkai and {Yang}, Jie and {Weathersby}, Stephen P. and {Park}, Suji and {Sie}, Edbert J. and {Jarillo-Herrero}, Pablo and {Fisher}, Ian R. and {Wang}, Xijie and {Demler}, Eugene and {Gedik}, Nuh},
	title = "{Role of Equilibrium Fluctuations in Light-Induced Order}",
	journal = {\prl},
	year = 2021,
	month = nov,
	volume = {127},
	number = {22},
	eid = {227401},
	pages = {227401},
	doi = {10.1103/PhysRevLett.127.227401},
	archivePrefix = {arXiv},
	eprint = {2110.00865},
	primaryClass = {cond-mat.mtrl-sci},
	adsurl = {https://ui.adsabs.harvard.edu/abs/2021PhRvL.127v7401Z}
}

@ARTICLE{Gedik20,
	author = {{Kogar}, Anshul and {Zong}, Alfred and {Dolgirev}, Pavel E. and {Shen}, Xiaozhe and {Straquadine}, Joshua and {Bie}, Ya-Qing and {Wang}, Xirui and {Rohwer}, Timm and {Tung}, I. -Cheng and {Yang}, Yafang and {Li}, Renkai and {Yang}, Jie and {Weathersby}, Stephen and {Park}, Suji and {Kozina}, Michael E. and {Sie}, Edbert J. and {Wen}, Haidan and {Jarillo-Herrero}, Pablo and {Fisher}, Ian R. and {Wang}, Xijie and {Gedik}, Nuh},
	title = "{Light-induced charge density wave in LaTe$_{3}$}",
	journal = {Nature Physics},
	year = 2020,
	month = feb,
	volume = {16},
	number = {2},
	pages = {159-163},
	doi = {10.1038/s41567-019-0705-3},
	archivePrefix = {arXiv},
	eprint = {1904.07472},
	primaryClass = {cond-mat.mtrl-sci},
	adsurl = {https://ui.adsabs.harvard.edu/abs/2020NatPh..16..159K}
}

@article{Mitrano16,
	author = {Mitrano, M. and Cantaluppi, A. and Nicoletti, D. and Kaiser, S. and Perucchi, A. and Lupi, S. and Di Pietro, P. and Pontiroli, D. and Ricc{\`o}, M. and Clark, S. R. and Jaksch, D. and Cavalleri, A.},
	date = {2016/02/01},
	doi = {10.1038/nature16522},
	id = {Mitrano2016},
	isbn = {1476-4687},
	journal = {Nature},
	number = {7591},
	pages = {461},
	title = "{Possible light-induced superconductivity in K$_3$C$_{60}$ at high temperature}",
	url = {https://doi.org/10.1038/nature16522},
	volume = {530},
	year = {2016}}

@ARTICLE{Millis20,
	author = {{Sun}, Zhiyuan and {Millis}, Andrew J.},
	title = "{Transient Trapping into Metastable States in Systems with Competing Orders}",
	journal = {Physical Review X},
	year = 2020,
	month = apr,
	volume = {10},
	number = {2},
	eid = {021028},
	pages = {021028},
	doi = {10.1103/PhysRevX.10.021028},
	archivePrefix = {arXiv},
	eprint = {1905.05341},
	primaryClass = {cond-mat.str-el},
	adsurl = {https://ui.adsabs.harvard.edu/abs/2020PhRvX..10b1028S}
}

@ARTICLE{Mitra21b,
	author = {{Natsheh}, Muath and {Gambassi}, Andrea and {Mitra}, Aditi},
	title = "{Critical properties of the prethermal Floquet time crystal}",
	journal = {Phys. Rev. B},
	year = 2021,
	month = jun,
	volume = {103},
	number = {22},
	eid = {224311},
	pages = {224311},
	doi = {10.1103/PhysRevB.103.224311},
	archivePrefix = {arXiv},
	eprint = {2103.10818},
	primaryClass = {cond-mat.stat-mech},
	adsurl = {https://ui.adsabs.harvard.edu/abs/2021PhRvB.103v4311N}
}

@ARTICLE{Mitra21a,
	author = {{Natsheh}, Muath and {Gambassi}, Andrea and {Mitra}, Aditi},
	title = "{Critical properties of the Floquet time crystal within the Gaussian approximation}",
	journal = {Phys. Rev. B},
	year = 2021,
	month = jan,
	volume = {103},
	number = {1},
	eid = {014305},
	pages = {014305},
	doi = {10.1103/PhysRevB.103.014305},
	archivePrefix = {arXiv},
	eprint = {2008.10560},
	primaryClass = {cond-mat.stat-mech},
	adsurl = {https://ui.adsabs.harvard.edu/abs/2021PhRvB.103a4305N}
}

@ARTICLE{Shimano23,
	author = {{Nishida}, Morihiko and {Katsumi}, Kota and {Song}, Dongjoon and {Eisaki}, Hiroshi and {Shimano}, Ryo},
	title = "{Light-induced coherent interlayer transport in stripe-ordered La$_{1.6 {\ensuremath{-}}x}$Nd$_{0.4}$Sr$_{x}$Cu O$_{4}$}",
	journal = {Phys. Rev. B},
	year = 2023,
	month = may,
	volume = {107},
	number = {17},
	eid = {174523},
	pages = {174523},
	doi = {10.1103/PhysRevB.107.174523},
	archivePrefix = {arXiv},
	eprint = {2303.01961},
	primaryClass = {cond-mat.supr-con},
	adsurl = {https://ui.adsabs.harvard.edu/abs/2023PhRvB.107q4523N}
}

@article{Shimano24,
	title = "{Emergence of light-induced superconducting-like state from the charge density wave state in high-${T}_{c}$ cuprate superconductors}",
	author = {Nishida, Morihiko and Song, Dongjoon and Hallas, Alannah M. and Eisaki, Hiroshi and Shimano, Ryo},
	journal = {Phys. Rev. B},
	volume = {110},
	issue = {22},
	pages = {224515},
	numpages = {12},
	year = {2024},
	month = {Dec},
	publisher = {American Physical Society},
	doi = {10.1103/PhysRevB.110.224515},
	url = {https://link.aps.org/doi/10.1103/PhysRevB.110.224515}
}

@ARTICLE{Chandran16,
	author = {{Chandran}, Anushya and {Sondhi}, S.~L.},
	title = "{Interaction-stabilized steady states in the driven O($N$) model}",
	journal = {Phys. Rev. B},
	year = 2016,
	month = may,
	volume = {93},
	number = {17},
	eid = {174305},
	pages = {174305},
	doi = {10.1103/PhysRevB.93.174305},
	archivePrefix = {arXiv},
	eprint = {1506.08836},
	primaryClass = {cond-mat.stat-mech},
	adsurl = {https://ui.adsabs.harvard.edu/abs/2016PhRvB..93q4305C}
}

@ARTICLE{Moessner25,
	author = {{Hou}, Yang and {Fu}, Zhanpeng and {Moessner}, Roderich and {Bukov}, Marin and {Zhao}, Hongzheng},
	title = "{Floquet-engineered emergent massive Nambu-Goldstone modes}",
	journal = {Phys. Rev. B},
	year = 2025,
	month = jul,
	volume = {112},
	number = {2},
	eid = {L020305},
	pages = {L020305},
	doi = {10.1103/9znd-dsm2},
	archivePrefix = {arXiv},
	eprint = {2409.01902},
	primaryClass = {quant-ph},
	adsurl = {https://ui.adsabs.harvard.edu/abs/2025PhRvB.112b0305H}
}

@ARTICLE{Cavalleri24,
	author = {{Fava}, S. and {De Vecchi}, G. and {Jotzu}, G. and {Buzzi}, M. and {Gebert}, T. and {Liu}, Y. and {Keimer}, B. and {Cavalleri}, A.},
	title = "{Magnetic field expulsion in optically driven YBa$_{2}$Cu$_{3}$O$_{6.48}$}",
	journal = {Nature},
	year = 2024,
	month = aug,
	volume = {632},
	number = {8023},
	pages = {75-80},
	doi = {10.1038/s41586-024-07635-2},
	archivePrefix = {arXiv},
	eprint = {2405.00848},
	primaryClass = {cond-mat.supr-con},
	adsurl = {https://ui.adsabs.harvard.edu/abs/2024Natur.632...75F}
}

@article{Carpene24,
	author = {Sayers, C. J. and Zhang, Y. and Sanders, C. E. and Chapman, R. T. and Wyatt, A. S. and Chatterjee, G. and Springate, E. and Cerullo, G. and Wolverson, D. and Da Como, E. and Carpene, E.},
	date = {2024/11/28},
	doi = {10.1038/s42005-024-01879-0},
	id = {Sayers2024},
	isbn = {2399-3650},
	journal = {Communications Physics},
	number = {1},
	pages = {389},
	title = {Mapping the nonequilibrium order parameter of a quasi-two dimensional charge density wave system},
	url = {https://doi.org/10.1038/s42005-024-01879-0},
	volume = {7},
	year = {2024}}

@ARTICLE{Komnik16,
	author = {{Komnik}, Andreas and {Thorwart}, Michael},
	title = "{BCS theory of driven superconductivity}",
	journal = {European Physical Journal B},
	year = 2016,
	month = nov,
	volume = {89},
	number = {11},
	eid = {244},
	pages = {244},
	doi = {10.1140/epjb/e2016-70528-1},
	archivePrefix = {arXiv},
	eprint = {1607.03858},
	primaryClass = {cond-mat.supr-con},
	adsurl = {https://ui.adsabs.harvard.edu/abs/2016EPJB...89..244K}
}

@ARTICLE{Babadi17,
	author = {{Babadi}, Mehrtash and {Knap}, Michael and {Martin}, Ivar and {Refael}, Gil and {Demler}, Eugene},
	title = "{Theory of parametrically amplified electron-phonon superconductivity}",
	journal = {Phys. Rev. B},
	year = 2017,
	month = jul,
	volume = {96},
	number = {1},
	eid = {014512},
	pages = {014512},
	doi = {10.1103/PhysRevB.96.014512},
	archivePrefix = {arXiv},
	eprint = {1702.02531},
	primaryClass = {cond-mat.supr-con},
	adsurl = {https://ui.adsabs.harvard.edu/abs/2017PhRvB..96a4512B}
}

@article{Wouters2007,
	title = {Excitations in a Nonequilibrium Bose-Einstein Condensate of Exciton Polaritons},
	author = {Wouters, Michiel and Carusotto, Iacopo},
	journal = {Phys. Rev. Lett.},
	volume = {99},
	issue = {14},
	pages = {140402},
	numpages = {4},
	year = {2007},
	month = {Oct},
	publisher = {American Physical Society},
	doi = {10.1103/PhysRevLett.99.140402},
	url = {https://link.aps.org/doi/10.1103/PhysRevLett.99.140402}
}

@article{Faraday1,
	title = {Parametric Excitation of a Bose-Einstein Condensate: From Faraday Waves to Granulation},
	author = {Nguyen, J. H. V. and Tsatsos, M. C. and Luo, D. and Lode, A. U. J. and Telles, G. D. and Bagnato, V. S. and Hulet, R. G.},
	journal = {Phys. Rev. X},
	volume = {9},
	issue = {1},
	pages = {011052},
	numpages = {11},
	year = {2019},
	month = {Mar},
	publisher = {American Physical Society},
	doi = {10.1103/PhysRevX.9.011052},
	url = {https://link.aps.org/doi/10.1103/PhysRevX.9.011052}
}

@article{Faraday2,
	title = {Observation of Pattern Stabilization in a Driven Superfluid},
	author = {Liebster, Nikolas and Sparn, Marius and Kath, Elinor and Duchene, Jelte and Fujii, Keisuke and G\"orlitz, Sarah L. and Enss, Tilman and Strobel, Helmut and Oberthaler, Markus K.},
	journal = {Phys. Rev. X},
	volume = {15},
	issue = {1},
	pages = {011026},
	numpages = {11},
	year = {2025},
	month = {Feb},
	publisher = {American Physical Society},
	doi = {10.1103/PhysRevX.15.011026},
	url = {https://link.aps.org/doi/10.1103/PhysRevX.15.011026}
}

@article{Faraday3,
	title={Supersolid-like sound modes in a driven quantum gas},
	author={Liebster, Nikolas and Sparn, Marius and Kath, Elinor and Duchene, Jelte and Strobel, Helmut and Oberthaler, Markus K},
	journal={Nature Physics},
	pages={1--7},
	year={2025},
	publisher={Nature Publishing Group UK London}
}

@article{hanai2019non,
	title={Non-Hermitian phase transition from a polariton Bose-Einstein condensate to a photon laser},
	author={Hanai, Ryo and Edelman, Alexander and Ohashi, Yoji and Littlewood, Peter B},
	journal={Physical review letters},
	volume={122},
	number={18},
	pages={185301},
	year={2019},
	publisher={APS}
}

@article{hanai2020critical,
	title={Critical fluctuations at a many-body exceptional point},
	author={Hanai, Ryo and Littlewood, Peter B},
	journal={Physical Review Research},
	volume={2},
	number={3},
	pages={033018},
	year={2020},
	publisher={APS}
}

@article{claude2025observation,
	title={Observation of the diffusive Nambu--Goldstone mode of a non-equilibrium phase transition},
	author={Claude, Ferdinand and Jacquet, Maxime J and Glorieux, Quentin and Wouters, Michiel and Giacobino, Elisabeth and Carusotto, Iacopo and Bramati, Alberto},
	journal={Nature Physics},
	pages={1--7},
	year={2025},
	publisher={Nature Publishing Group UK London}
}

@article{heiss2012physics,
	title={The physics of exceptional points},
	author={Heiss, Walter D},
	journal={Journal of Physics A: Mathematical and Theoretical},
	volume={45},
	number={44},
	pages={444016},
	year={2012},
	publisher={IOP Publishing}
}

@inproceedings{berges2004introduction,
	title={Introduction to nonequilibrium quantum field theory},
	author={Berges, J{\"u}rgen},
	booktitle={AIP Conf. Proc.},
	volume={739},
	number={1},
	pages={3--62},
	year={2004},
	organization={AIP}
}

@article{Hill,
	author = {Hill, G.},
	title = {On the part of the motion of the lunar perigee which is a function of the mean motions of the sun and moon},
	journal = {Acta. Math.},
	volume = {8},
	year = {1886},
	pages = {1}
}

@article{Mathieu,
	author = {Mathieu, E.},
	title = {Mémoire sur le mouvement vibratoire d’une membrane de forme elliptique},
	journal = {Journal de Mathématiques Pures et Appliquées},
	volume = {13},
	year = {1868},
	pages = {137-203}
}

@article{Floquet,
	author = {Floquet, G.},
	title = {Sur les équations différentielles linéaires à coefficients périodiques},
	journal = {Annales Scientifiques de l’École Normale Supérieure},
	volume = {12},
	year = {1883},
	pages = {47-88}
}

@article{Dai2021,
	title = {Superconductinglike response in a driven gapped bosonic system},
	author = {Dai, Zhehao and Lee, Patrick A.},
	journal = {Phys. Rev. B},
	volume = {104},
	issue = {5},
	pages = {054512},
	numpages = {15},
	year = {2021},
	month = {Aug},
	publisher = {American Physical Society},
	doi = {10.1103/PhysRevB.104.054512},
	url = {https://link.aps.org/doi/10.1103/PhysRevB.104.054512}
}

@ARTICLE{Natseh2021_2,
	author = {{Natsheh}, Muath and {Gambassi}, Andrea and {Mitra}, Aditi},
	title = "{Critical properties of the prethermal Floquet time crystal}",
	journal = {\prb},
	year = 2021,
	month = jun,
	volume = {103},
	number = {22},
	eid = {224311},
	pages = {224311},
	doi = {10.1103/PhysRevB.103.224311},
	archivePrefix = {arXiv},
	eprint = {2103.10818},
	primaryClass = {cond-mat.stat-mech},
	adsurl = {https://ui.adsabs.harvard.edu/abs/2021PhRvB.103v4311N}
}

@ARTICLE{Chandran2016,
	author = {{Chandran}, Anushya and {Sondhi}, S.~L.},
	title = "{Interaction-stabilized steady states in the driven O (N ) model}",
	journal = {\prb},
	year = 2016,
	month = may,
	volume = {93},
	number = {17},
	eid = {174305},
	pages = {174305},
	doi = {10.1103/PhysRevB.93.174305},
	archivePrefix = {arXiv},
	eprint = {1506.08836},
	primaryClass = {cond-mat.stat-mech},
	adsurl = {https://ui.adsabs.harvard.edu/abs/2016PhRvB..93q4305C}
}

@book{kamenev2023Book,
	title        = {Field Theory of Non-Equilibrium Systems},
	author       = {Kamenev, Alex},
	year         = {2023},
	edition      = {2},
	publisher    = {Cambridge University Press},
	address      = {Cambridge},
	isbn         = {9781108488259}
}

@book{Tinkham2004,
	author    = {Michael Tinkham},
	title     = {Introduction to Superconductivity},
	edition   = {2nd},
	publisher = {Dover Publications},
	year      = {2004},
	isbn      = {9780486435039},
	pages     = {454},
}

@misc{MathieuNote,
	note={We use the convention that the cosine-Mathieu functions of the first kind $\mathrm{ce}_{n}(q,\tau)$ obey the equation $[\partial_\tau^2+a_{n}(q)-2q\cos(2\tau)]\mathrm{ce}_{n}(q,\tau)=0$, $\partial_\tau\mathrm{ce}_{n}(q,\tau)|_{\tau=0}=0$, with $a_{n}(q)$ the type-A Mathieu characteristic numbers. Here we choose a normalization such that $\frac{1}{2\pi}\int_0^{2\pi}\!d\tau\,[\mathrm{ce}_{n}(q,\tau)]^2=\frac{1}{2}$. Note that the odd-$n$ functions are $2\pi$-periodic, whereas the even ones are $\pi$-periodic.}
}

@article{Nova2019,
	author       = {Nova, Tobia F. and Cartella, Anton and Cantaluppi, Andrea and Jakob, Gerhard and Macke, Stefan and Poccia, Niels and Bruhacs, Attila and Radek, Petr and Gaal, Robert and Le Toullec, Romain and Olivier, Lionel and Vedmedenko, Egid Yu. and Rini, Matteo and Cavalleri, Andrea},
	title        = {Metastable ferroelectricity in optically strained $ \mathrm{SrTiO}_{3} $},
	journal      = {Science},
	volume       = {364},
	number       = {6442},
	pages        = {1075--1079},
	year         = {2019},
	doi          = {10.1126/science.aaw4911}
}

@ARTICLE{Padma2025,
	author = {{Padma}, Hari and {Sharma}, Prakash and {TenHuisen}, Sophia F.~R. and {Glerean}, Filippo and {Roll}, Antoine and {Zhou}, Pan and {Kundu}, Sarbajaya and {Romaguera}, Arnau and {Skoropata}, Elizabeth and {Ueda}, Hiroki and {Liu}, Biaolong and {Paris}, Eugenio and {Wang}, Yu and {Huat Lee}, Seng and {Mao}, Zhiqiang and {Dean}, Mark P.~M. and {Huang}, Edwin W. and {Razzoli}, Elia and {Wang}, Yao and {Mitrano}, Matteo},
	title = "{A light-induced charge order mode in a metastable cuprate ladder}",
	journal = {arXiv e-prints},
	year = 2025,
	month = oct,
	eid = {arXiv:2510.24686},
	pages = {arXiv:2510.24686},
	doi = {10.48550/arXiv.2510.24686},
	archivePrefix = {arXiv},
	eprint = {2510.24686},
	primaryClass = {cond-mat.str-el},
	adsurl = {https://ui.adsabs.harvard.edu/abs/2025arXiv251024686P}
}

@article{Fechner2024,
	author = {{Fechner}, M. and {F{\"o}rst}, M. and {Orenstein}, G. and {Krapivin}, V. and {Disa}, A.~S. and {Buzzi}, M. and {von Hoegen}, A. and {de la Pena}, G. and {Nguyen}, Q.~L. and {Mankowsky}, R. and {Sander}, M. and {Lemke}, H. and {Deng}, Y. and {Trigo}, M. and {Cavalleri}, A.},
	title = "{Quenched lattice fluctuations in optically driven SrTiO$_{3}$}",
	journal = {Nature Materials},
	year = 2024,
	month = mar,
	volume = {23},
	number = {3},
	pages = {363-368},
	doi = {10.1038/s41563-023-01791-y},
	archivePrefix = {arXiv},
	eprint = {2301.08703},
	primaryClass = {cond-mat.mtrl-sci},
	adsurl = {https://ui.adsabs.harvard.edu/abs/2024NatMa..23..363F}
}

@ARTICLE{Kovacic2018,
	author = {{Kovacic}, Ivana and {Rand}, Richard and {Mohamed Sah}, Si},
	title = "{Mathieu's Equation and Its Generalizations: Overview of Stability Charts and Their Features}",
	journal = {Applied Mechanics Reviews},
	year = 2018,
	month = feb,
	volume = {70},
	number = {2},
	pages = {020802},
	doi = {10.1115/1.4039144},
	adsurl = {https://ui.adsabs.harvard.edu/abs/2018ApMRv..70b0802K}
}

@book{Strogatz2018,
	title={Nonlinear Dynamics and Chaos: With Applications to Physics, Biology, Chemistry, and Engineering},
	author={Strogatz, Steven H.},
	year={2018},
	publisher={CRC Press},
	edition={2nd},
	isbn={978-0813350848}
}

@book{Nayfeh1979,
	title={Nonlinear Oscillations},
	author={Nayfeh, Ali H. and Mook, Dean T.},
	year={1979},
	publisher={John Wiley \& Sons},
	isbn={978-0471033790}
}

@article{Arnold1965,
	title={Small denominators. I. Mappings of the circumference onto itself},
	author={Arnold, V. I.},
	journal={Izvestiya Akademii Nauk SSSR, Seriya Matematicheskaya},
	volume={25},
	number={1},
	pages={21--86},
	year={1965}
}

@article{Michael2025,
	author       = {M. H. Michael and E. Demler and P. Lee},
	title        = {Parametrically amplified Josephson plasma waves in $\mathrm{YBa_2Cu_3O_{6+x}}$: evidence for local superconducting fluctuations up to the pseudogap temperature $T^*$},
	journal      = {arXiv:2505.03358 [cond-mat.supr-con]},
	year         = {2025},
	doi          = {10.48550/arXiv.2505.03358}
}

@misc{Diessel_inprep,
	author = {{Diessel}, Oriana K. and {Bonetti}, Pietro M. and {Sachdev}, Subir},
	note={{\it in preparation}}
}

@article{Michael2024,
	author       = {M. H. Michael and D. De Santis and E. A. Demler and P. A. Lee},
	title        = {Giant dynamical paramagnetism in the driven pseudogap phase of $\mathrm{YBa_2Cu_3O_{6+x}}$},
	journal      = {arXiv:2410.12919 [cond-mat.supr-con]},
	year         = {2024},
	doi          = {10.48550/arXiv.2410.12919}
}

@article{Kaplan2025b,
	author       = {D. Kaplan and P. Volkov and A. Cavalleri and P. Chandra},
	title        = {Optically-Induced Faraday-Goldstone Waves},
	journal      = {arXiv:2511.07320 [cond-mat.mes-hall]},
	year         = {2025},
	doi          = {10.48550/arXiv.2511.07320}
}

@ARTICLE{Zelle2024,
	author = {{Zelle}, Carl Philipp and {Daviet}, Romain and {Rosch}, Achim and {Diehl}, Sebastian},
	title = "{Universal Phenomenology at Critical Exceptional Points of Nonequilibrium O (N ) Models}",
	journal = {Physical Review X},
	year = 2024,
	month = jun,
	volume = {14},
	number = {2},
	eid = {021052},
	pages = {021052},
	doi = {10.1103/PhysRevX.14.021052},
	archivePrefix = {arXiv},
	eprint = {2304.09207},
	primaryClass = {cond-mat.stat-mech},
	adsurl = {https://ui.adsabs.harvard.edu/abs/2024PhRvX..14b1052Z}
}

@ARTICLE{Daviet2024,
	author = {{Daviet}, Romain and {Zelle}, Carl Philipp and {Rosch}, Achim and {Diehl}, Sebastian},
	title = "{Nonequilibrium Criticality at the Onset of Time-Crystalline Order}",
	journal = {\prl},
	year = 2024,
	month = apr,
	volume = {132},
	number = {16},
	eid = {167102},
	pages = {167102},
	doi = {10.1103/PhysRevLett.132.167102},
	archivePrefix = {arXiv},
	eprint = {2312.13372},
	primaryClass = {cond-mat.stat-mech},
	adsurl = {https://ui.adsabs.harvard.edu/abs/2024PhRvL.132p7102D}
}

@ARTICLE{Weidinger2017,
	author = {{Weidinger}, Simon A. and {Knap}, Michael},
	title = "{Floquet prethermalization and regimes of heating in a periodically driven, interacting quantum system}",
	journal = {Scientific Reports},
	year = 2017,
	month = apr,
	volume = {7},
	eid = {45382},
	pages = {45382},
	doi = {10.1038/srep45382},
	archivePrefix = {arXiv},
	eprint = {1609.09089},
	primaryClass = {cond-mat.quant-gas},
	adsurl = {https://ui.adsabs.harvard.edu/abs/2017NatSR...745382W}
}

@ARTICLE{Knap2016,
	author = {{Knap}, Michael and {Babadi}, Mehrtash and {Refael}, Gil and {Martin}, Ivar and {Demler}, Eugene},
	title = "{Dynamical Cooper pairing in nonequilibrium electron-phonon systems}",
	journal = {\prb},
	year = 2016,
	month = dec,
	volume = {94},
	number = {21},
	eid = {214504},
	pages = {214504},
	doi = {10.1103/PhysRevB.94.214504},
	archivePrefix = {arXiv},
	eprint = {1511.07874},
	primaryClass = {cond-mat.supr-con},
	adsurl = {https://ui.adsabs.harvard.edu/abs/2016PhRvB..94u4504K}
}

@ARTICLE{Sieberer2025,
	author = {{Sieberer}, Lukas M. and {Buchhold}, Michael and {Marino}, Jamir and {Diehl}, Sebastian},
	title = "{Universality in driven open quantum matter}",
	journal = {Reviews of Modern Physics},
	year = 2025,
	month = apr,
	volume = {97},
	number = {2},
	eid = {025004},
	pages = {025004},
	doi = {10.1103/RevModPhys.97.025004},
	archivePrefix = {arXiv},
	eprint = {2312.03073},
	primaryClass = {cond-mat.stat-mech},
	adsurl = {https://ui.adsabs.harvard.edu/abs/2025RvMP...97b5004S}
}

@ARTICLE{Eckhardt2024,
	author = {{Eckhardt}, Christian J. and {Chattopadhyay}, Sambuddha and {Kennes}, Dante M. and {Demler}, Eugene A. and {Sentef}, Michael A. and {Michael}, Marios H.},
	title = "{Theory of resonantly enhanced photo-induced superconductivity}",
	journal = {Nature Communications},
	year = 2024,
	month = mar,
	volume = {15},
	eid = {2300},
	pages = {2300},
	doi = {10.1038/s41467-024-46632-x},
	archivePrefix = {arXiv},
	eprint = {2303.02176},
	primaryClass = {cond-mat.supr-con},
	adsurl = {https://ui.adsabs.harvard.edu/abs/2024NatCo..15.2300E}
}

@ARTICLE{Kennes2017,
	author = {{Kennes}, Dante M. and {Wilner}, Eli Y. and {Reichman}, David R. and {Millis}, Andrew J.},
	title = "{Transient superconductivity from electronic squeezing of optically pumped phonons}",
	journal = {Nature Physics},
	year = 2017,
	month = may,
	volume = {13},
	number = {5},
	pages = {479-483},
	doi = {10.1038/nphys4024},
	archivePrefix = {arXiv},
	eprint = {1609.03802},
	primaryClass = {cond-mat.supr-con},
	adsurl = {https://ui.adsabs.harvard.edu/abs/2017NatPh..13..479K}
}

@ARTICLE{Homann2020,
	author = {{Homann}, Guido and {Cosme}, Jayson G. and {Mathey}, Ludwig},
	title = "{Higgs time crystal in a high-T$_{c}$ superconductor}",
	journal = {Physical Review Research},
	year = 2020,
	month = nov,
	volume = {2},
	number = {4},
	eid = {043214},
	pages = {043214},
	doi = {10.1103/PhysRevResearch.2.043214},
	archivePrefix = {arXiv},
	eprint = {2004.13383},
	primaryClass = {cond-mat.supr-con},
	adsurl = {https://ui.adsabs.harvard.edu/abs/2020PhRvR...2d3214H}
}
\appendix

\section{Hubbard-Stratonovich transformation and saddle point}
\label{app:HST and SP}
In this section we present a derivation of the large-$N$ equations used in the main text.
We start by writing the Lagrangian Eq.~\eqref{eq: b potential} on the Keldysh double time contour and we add a dissipative term~\cite{kamenev2023Book, Diehl16,Sieberer2025}.  We also define $b_\cl$ and $b_\qn$ as the classical and quantum components of $b$, respectively.
We then perform a Hubbard-Stratonovich transformation to decouple the interaction term to get the Lagrangian density

\begin{subequations}
	\begin{align}
		\mathcal{L}_0=&(\partial_\mu\boldsymbol{b})^\dagger\kappa^1\,(\partial_\mu\boldsymbol{b})\nonumber\\
		&-\gamma \boldsymbol{b}^\dagger (-i\kappa^2)\partial_t \boldsymbol{b}+4i\gamma T |b_\qn|^2\,,\\
		\mathcal{L}_\mathrm{HS}=&\mathcal{L}_0 
		+\frac{4N}{u}\lambda_q\lambda_c \nonumber
		\\&- \lambda_\cl 
		\left(|b_c|^2+|b_q|^2\right)\nonumber\\
		&-\lambda_\qn\left(b^\dagger_q b_c + \text{c.c.}\right)\,,
	\end{align}
\end{subequations}
where the matrices $\kappa^\alpha$ are Pauli matrices acting in cl-q space, with $\kappa^0=\mathbbm{1}_2$.
At this point, we can write $\boldsymbol{b}=\sqrt{2N}\bar{\boldsymbol{b}}+\delta\boldsymbol{b}$, with $\sqrt{2N}\bar{\boldsymbol{b}}=\langle\boldsymbol{b}\rangle$, and integrate out $\delta\boldsymbol{b}$ to get the effective action
\begin{equation}
	\label{eq: effective action large N}
	\begin{split}
		\mathcal{S}_\mathrm{eff}[\bar{\boldsymbol{b}},\boldsymbol{\lambda}]=&\int_{\rr,t}\,\mathcal{L}_\mathrm{HS}[\sqrt{2N}\bar{\boldsymbol{b}}]\\
		&+2N\mathrm{TrLog}\left(i\mathcal{G}^{-1}[\boldsymbol{\lambda}]\right)\,,
	\end{split}
\end{equation}
where we have defined 
\begin{equation}
	\begin{split}
		\mathcal{G}^{-1}[\boldsymbol{\lambda}]=&-\partial_\mu \partial^\mu\kappa^1-\gamma(-i\kappa^2)\partial_t+2i \gamma T(\kappa^0-\kappa^3)\\
		&-\lambda_\cl \,\kappa^1-\lambda_\qn\,\kappa^0\,,
	\end{split}
\end{equation}
as a $2\times2$ matrix acting on the $\cl$-$\qn$ index. One can see by inspection in that both terms in Eq.~\eqref{eq: effective action large N} the dependence on $N$ is a simple multiplicative term. Taking the $N\to\infty$ limit makes the saddle point approximation exact, so we can study the equations of motion for the $\boldsymbol{X}=\{\bar{\boldsymbol{b}},\boldsymbol{\lambda}\}$ fields:
\begin{equation}
	\frac{\delta [\mathcal{S}_\mathrm{eff}/(2N)]}{\delta X_\qn}\bigg\rvert_{X_\qn=0}=0\,.
\end{equation}
We now assume that the condensates of the fields $X(t)\equiv X_\cl$ depend on time but not on space. We further impose $\bar{b}_\cl=(\bar{b},0,\dots,0)$. We get:
\begin{subequations}\label{eq:large N eqs1}
	\begin{align}
		& \lambda(t)=r(t)+u\left[\bar{b}^2(t)+\int_\qq C(q,t)\right],\\
		&\left[\partial_t^2+\gamma\partial_t + \lambda(t)\right]\bar{b}(t)=0\,,
	\end{align}
\end{subequations}
Where we have defined $C(q,t)=iG^K_N(|\bq|,t,t)$, with $G^K_N(\bq,t,t')$ the Keldysh ($\cl$-$\cl$) component of the inverse of the operator $\mathcal{G}^{-1}[\boldsymbol{\lambda}=(\lambda(t),0)]$.

The evolution equations for the large-$N$ Green's function are obtained by inverting the operator $\mathcal{G}^{-1}[\boldsymbol{\lambda}=(\lambda(t),0)]$:
%
\begin{subequations}\label{eq: time evolution Gfs}
	\begin{align}
		&\left[\partial_{t_1}^2 +|\bq|^2 +\gamma\partial_{t_1} + \lambda(t_1)\right]G^K_N(\bq,t_1,t_2)=\nonumber\\
		&\hskip4cm 4i\gamma T G^A_N(\bq,t_1,t_2)\,,\label{eq: evolution GK}\\
		-&\left[\partial_{t_1}^2 +|\bq|^2 +\gamma\partial_{t_1} + \lambda(t_1)\right]G^R_N(\bq,t_1,t_2) =\nonumber\\ 
		&\hskip5cm\delta(t_1-t_2)\,,\label{eq: evolution GR}\\
		&G^A_N(\bq,t_1,t_2)=\left[G^R_N(\bq,t_2,t_1)\right]^*\,.
	\end{align}
\end{subequations}
%
We now additionally define $D(q,t)=\partial_{t_1}\partial_{t_2}G^K_N(t_1,t_2)\rvert_{t_1=t_2=t}$ and note that
\begin{subequations}
	\begin{align}
		&\partial_t C(q,t)=2i\partial_{t_1} G^K_N(|\bq|,t_1,t_2)\rvert_{t_1=t_2=t}\,,\\
		&\partial_t^2 C(q,t)=2i\partial^2_{t_1} G^K_N(|\bq|,t_1,t_2)\rvert_{t_1=t_2=t} + 2D(q,t)\,,\\
		&\partial_t D(q,t)=2i\partial^2_{t_1}\partial_{t_2} G^K_N(|\bq|,t_1,t_2)\rvert_{t_1=t_2=t}
	\end{align}
\end{subequations}
Setting $t_1=t_2=t$ in Eq.~\eqref{eq: evolution GK}, multiplying it by $i$ on both hand sides, and using the above relations we get
\begin{multline}
	\label{eq: C eq appendix}
	\frac{1}{2}\partial_t^2C(q,t)-D(q,t)+\frac{1}{2}\gamma\, \partial_t C(q,t)
	\\+[q^2+\lambda(t)]C(q,t)=-4\gamma T G^A_N(|\bq|,t,t).
\end{multline}
Deriving Eq.~\eqref{eq: evolution GK} with respect to $t_2$, multiplying by $i$, and setting $t_1=t_2=t$, we obtain
\begin{multline}\label{eq: D eq appendix}
	\frac{1}{2}D(q,t)+\gamma\hat{D}(q,t)+\frac{1}{2}[q^2+\lambda(t)]\partial_t C(q,t)\\=-4\gamma T\,\partial_{t_2} G^A_N(|\bq|,t_1,t_2)\rvert_{t_1=t_2=t}\,.
\end{multline}
Recalling that 
\begin{subequations}
	\begin{align}
		&G^A(\bq,t_1,t_2)=\frac{1}{2}\theta(t_2-t_1)\langle[b^\dagger_\bq(t),b_\bq(t')]\rangle \,,\\
		&\partial_{t_2} G^A(\bq,t_1,t_2)=\frac{1}{2}\theta(t_2-t_1)\langle[b^\dagger_\bq(t),\Pi_\bq(t')]\rangle \,,
	\end{align}
\end{subequations}
with $\theta(0)=1/2$ and $\Pi_q(t)$ the canonically conjugate operator to $b^\dagger_\bq$, and noting that our model requires $[b^\dagger,b]=0$ and $[\Pi,b^\dagger]=1$, we deduce that the right hand sides of Eqs.~\eqref{eq: C eq appendix} and \eqref{eq: D eq appendix} are 0 and $\gamma T$, respectively. In this way, we get Eqs.~\eqref{eq:EoM2}, \eqref{eq:EoM3}.

\section{Expressions for the coefficients $C_{n}(q)$ in the steady state}
\label{app:C_Floquet}
Combining Eqs.~\eqref{eq:EoM}, we get
\begin{equation}
	\begin{split}
		\Big\{
		&\partial_t^3+3\gamma\partial_t^2+\left[4q^2+4\lambda(t)+2\gamma^2\right]\partial_t\\
		&+\left[4\gamma q^2+4\gamma \lambda_\pm(t)+2\dot{\lambda}(t)\right]
		\Big\}C(q,t)=4\gamma T\,.
	\end{split}
\end{equation}
Expanding $C(q,t)=\sum_{n=-\infty}^{+\infty}C_{n}(q)\,e^{-in\Omega t}$, we get $C_n(q) = 4\gamma T \mathcal{C}^{-1}_{n0}(q)$, with
\begin{equation}\label{eq:C_correlator_Floquet}
	\begin{split}
		\mathcal{C}_{nm}(q)=&\bigg[(-in\Omega)^3+3\gamma(-in\Omega)^2\\ &+(4q^2+2\gamma^2)(-in\Omega)+
		4\gamma q^2\bigg]\delta_{nm} \\ 
		&\bigg[ 4(-im\Omega) + 4\gamma + 2 (-i(n-m)\Omega)\bigg] \lambda_{n-m}\,.
	\end{split}
\end{equation}
%

\section{Iterative solution of Floquet equations}
\label{app:truncation}
We can iteratively look for nontrivial steady states using the following algorithm
\begin{enumerate}
	\itemsep 0pt
	\item Start with a guess for $\lambda_n$ (e.g. $R_n=r_n$),
	\item Update $\lambda_0$ by imposing $\mathrm{det}\mathcal{K}(\nu)=0$ at $\omega=0$ or $\Omega/2$,
	\item Calculate $\bar{b}_n$ up to a constant $\mathcal{B}$ from the kernel of $\mathcal{K}(\nu)$ ($\bar{b}_n=\mathcal{B}\,\hat{b}_n$ with $\sum_n|\hat{b}_n|^2=1$),
	\item Calculate the global constant $\mathcal{B}$ at $n=0$, using Eq.~\eqref{eq:C_correlator_Floquet} for $C_n(q)$,
	\item Update $\lambda_{|n|\geq 1}$ from Eq.~\eqref{eq: lambda equation floquet} at $|n|\geq 1$,
	\item repeat 1.-5. until convergence is reached.
\end{enumerate}
If the algorithm returns a nonzero value of $\mathcal{B}$, then a nontrivial oscillating symmetry broken steady state is induced by the interactions. 
\section{Determination of finite-$q$ instabilities}
\label{app: finite-q}
In this section, we discuss the method we employed to detect finite-$q$ instabilities. This method relies on the calculation of the Floquet band structures in the steady state. The retarded Green's function in Sambe space reads 
\begin{equation}
	\begin{split}
		\left[G^R_{nm}(\bq,\omega)\right]^{-1} = &\left[\left(\omega+n\Omega+i\frac{\gamma}{2}\right)^2
		-\tilde{\epsilon}_\bq^2\right]\delta_{nm}\\ &\hskip2cm- \sum_{\ell\neq 0} \lambda_\ell\, \delta_{n,m+\ell}\,,
	\end{split}
\end{equation}
with $\tilde{\epsilon}_\bq=\sqrt{q^2+\lambda_0-\gamma^2/4}$, and $\lambda_\ell$ are calculated numerically by solving the equations in real time and Fourier transforming $\lambda(t)$ in the steady state. 
The Floquet quasi-bands $E_n(\bq)$ are defined as the poles of $G^R_{nm}(\bq,\omega)$ for complex $\omega$. We can numerically calculate $E_n(\bq)$ by calculating the eigenvalues $\tilde{E}_n(\bq)$ of the matrix
\begin{equation}
	M_\bq = \left(
	\begin{array}{c|c}
		0 & \left[\tilde{\epsilon}_\bq^2-n^2\Omega^2\right]\delta_{nm} + \sum_{\ell\neq 1}\lambda_\ell \,\delta_{n,m+\ell} \\\hline
		\mathbbm{1} & 2 n\Omega\,\delta_{nm} 
	\end{array}
	\right)\,.
\end{equation}
While each one of the blocks is an infinite matrix, we truncate them to $N\times N$ matrices to make numerical calculations amenable. In the phase diagram in Fig.~\ref{Fig:PhaseDiagram} we used $N=11$.
The Floquet quasi-bands are given by $E_n(\bq)=\tilde{E}_n(\bq)-i\gamma/2$. Note that $E_n(\bq)$ obeys $E_{n+m}(\bq)=E_n(\bq)+m\Omega$. To detect an instability, we focus on the imaginary part of $E_n(\bq)$. Due to causality, the poles of the retarded Green's function should be located only in the lower complex half plane. Whenever the imaginary part of $E_n(\bq)$ turns zero for some $\bq=\bQ$ or becomes even positive, we say that the system has an instability. If this occurs at a nonzero value of $\bQ$, then we have a finite momentum instability. To distinguish between period-preserving and period-doubling instabilities, we look at $[\mathrm{Re}\, E_n(\bQ)]\,\mathrm{mod}\,\Omega$. If this quantity equals 0 or $\Omega$, then the instability is period-preserving, if instead it equals $\Omega/2$, then it is period-doubling. Note that these are the only possibilities. Typical behaviors of the imaginary parts of $E_n(\bq)$ are shown in the right column of Fig.~\ref{Fig:Dispersions}, where a 2$\times$2 truncation was performed. Despite the small size of the truncated Green's function, the qualitative behavior of $\mathrm{Im}\, E_n(\bq)$ is captured correctly.

\begin{widetext}
	\section{Electromagnetic response}
	\label{app: electromagnetic response}
	In this Appendix, we evaluate the electromagnetic response of the O($N$) theory discussed in the main text in the large-$N$ limit. In the following, we assume $N$ to be even and construct the $N/2$-dimensional complex vector 
	\begin{equation}
		\phi=\frac{1}{\sqrt{2}}\left(
		\begin{array}{c}
			b_1+ib_{N/2+1} \\ 
			b_2+ib_{N/2+2} \\
			\vdots\\
			b_{N/2-1}+ib_{N-1}\\
			b_{N/2}+ib_{N}
		\end{array}
		\right)\,.
	\end{equation} 
	In this way, we can rewrite the Lagrangian as
	\begin{equation}
		\mathcal{L}=|\partial_\mu \phi|^2-r|\phi|^2-\frac{u}{N}(|\phi|^2)^2\,.
	\end{equation}
	The coupling to the electromagnetic field is then realized by promoting the spatio-temporal derivative $\partial_\mu$ to a \textit{covariant} derivative $D_\mu=\partial_\mu + i e_\star A_\mu$. For simplicity, we will set $e_\star=1$ and reinstate it when needed using dimensional analysis. In essence, coupling the theory to the electromagnetic field breaks its symmetry group O($N$) down to U($N/2$)=U(1)$\times$ SU($N/2$), where the U(1) subgroup of U($N/2$) is gauged and the remaining SU($N/2$) is a global symmetry. As we are ultimately interested in the $N=2$ case, this feature becomes irrelevant. 
	\subsection{Electromagnetic response at large-$N$}
	At large-$N$, the partition function can be written as 
	\begin{equation}
		\int\mathcal{D}\vec{X} e^{iN\mathcal{S}[\vec{X};A_\mu]}\,,
	\end{equation}
	where $A_\mu$ is an external electromagnetic vector potential that we use to probe the system. For the O($N$) theory we have $\vec{X}=\{\bar{\boldsymbol{b}},\boldsymbol{\lambda}\}$ and $\mathcal{S}[\vec{X},A_\mu=0]$ is given by Eq.~\eqref{eq: effective action large N}. The coupling to the electromagnetic field occurs via the covariant derivative $D_\mu=\partial_\mu\mathbbm{1}+ i A_{\cl,\mu}\mathbbm{1}\otimes\kappa^0+iA_{\qn,\mu}\mathbbm{1}\otimes\kappa^1$.
	
	Because we are interested in linear response, we expand $\mathcal{S}[\vec{X};A_\mu]$ up to second order in $A_\mu$:
	\begin{equation}
		\mathcal{S}[\vec{X};A_\mu]\approx \mathcal{S}_0[\vec{X}]+\left(\mathcal{S}_1^\mu[\vec{X}],A_\mu\right)+\frac{1}{2} \left(A_\mu,\mathcal{S}_2^{\mu\nu}[\vec{X}] A_\nu\right)\,,
	\end{equation}
	where $(\bullet,\bullet)$ is a shorthand for a sum over all internal indices. We now expand the saddle point solution $\vec{X}^*[A_\mu]$ in powers of $A_\mu$:
	\begin{equation}
		\vec{X}^*[A_\mu]=\vec{X}^*_0+\left(\vec{X}_1^{*\mu},A_\mu\right)+\frac{1}{2}\left(A_\mu,\vec{X}_2^{*\mu\nu}A_\nu\right)\,.
	\end{equation}
	The saddle point action is given by 
	\begin{equation}
		\mathcal{S}^*=\mathcal{S}_0[\vec{X}^*_0]+\left(\mathcal{S}^\mu_1[\vec{X}^*_0],A_\mu\right)+\frac{1}{2}\left(A_\mu,\left[\mathcal{S}^{\mu\nu}_2[\vec{X}^*_0]-\Gamma_1^\mu H_0^{-1}\Gamma_1^\nu\right]A_\nu\right)\,,
	\end{equation}
	with $\Gamma_1^\mu=\delta_X \mathcal{S}_1^\mu[\vec{X}^*_0]$ and $H_0=\delta_X^2\mathcal{S}_0[\vec{X}^*_0]$. This tells us that the electromagnetic kernel is given by the tensor $\mathcal{S}^{\mu\nu}_2[\vec{X}^*_0]-\Gamma_1^\mu H_0^{-1}\Gamma_1^\nu$. In particular, we are interested in the spatial components of the $\bq=0$ kernel $K^{\alpha\beta}(t,t')\sim \langle A_{\qn,\alpha}(q=0,t) A_{\cl,\beta}(q=0,t')\rangle$, where $\alpha$ and $\beta$ are only spatial indices. For the special large-$N$ solutions discussed in the main text, the electromagnetic Kernel takes the form:
	\begin{equation}
		\begin{split}
			K^{\alpha\beta}(t,t')=&\bar{b}_{\bQ}(t)\bar{b}_{-\bQ}(t)\delta_{t,t'}\delta_{\alpha\beta}+\int_{\bq} \mathrm{Tr}\left[iG^K_N(q;t,t)\right]\delta_{t,t'}\delta_{\alpha\beta}\nonumber\\
			&+i\int_\bq 2q_\alpha q_\beta\mathrm{Tr}\left[G^R_N(q;t,t')G_N^K(q;t',t)+G^K_N(q;t,t')G_N^A(q;t',t)\right] \label{eq: EM Kernel0}\\
			&-\lim_{q\to 0} q_\alpha q_\beta\,\,\mathrm{Tr}\left[G_N^R(q;t,t')\right]\,\bar{b}_{\bQ}(t)\bar{b}_{-\bQ}(t')\nonumber\,,
		\end{split}
	\end{equation}
	here $G_N^{R|A|K}(q;t,t')$ are the retarded, advanced and Keldysh components of the Green's function, whose time evolution is governed by equations~\eqref{eq: time evolution Gfs}. Here, $\int_\bq$ is a shorthand for $\int_{q\in[\Lambda_\mathrm{IR},\Lambda_\mathrm{UV}]}\!\frac{d^2\bq}{(2\pi)^2}$. 
	\subsection{Meissner effect and optical conductivity}
	We now focus on the optical conductivity and the Meissner effect. In the steady state, we can represent the Green's function as an infinite matrix using Floquet theory.
	\begin{subequations}
		\begin{align}
			&\left[\boldsymbol{G}^{R|A}(q,\omega)\right]^{-1}_{nm}=\left[(\omega+n\Omega)^2+i\gamma(\omega+n\Omega)-q^2\right]\delta_{nm}-\sum_{\ell=-\infty}^{+\infty}\lambda_\ell\, \delta_{n,m+\ell}\,,\label{eq: GRA Floquet}\\
			&\boldsymbol{G}^K(q,\omega)=-2i\gamma \boldsymbol{G}^R(q,\omega) \boldsymbol{F}(\omega) \boldsymbol{G}^A(q,\omega)\,,\label{eq: GK Floquet}\\
			&[\boldsymbol{F}(\omega)]_{mn}=(\omega+n\Omega) n_B(\omega+n\Omega)\,\delta_{mn}\,,
		\end{align}
	\end{subequations}
	with $n_B(x)=1/(e^{x/T}-1)$ the Bose-Einstein distribution function. In the classical (high-temperature) limit considered in the main text, the distribution function becomes $\boldsymbol{F}(\omega)=2T\delta_{mn}$. For simplicity, we work in the Coulomb gauge, $\vec{\nabla}\cdot \vec{A}=0$, which allows us to neglect the last term in Eq.~\eqref{eq: EM Kernel0}. In general, those terms will ensure that the electromagnetic kernel remains gauge invariant. In the steady state, we can expand the kernel in Floquet harmonics as
	\begin{equation}
		K^{\alpha\beta}(t,t')=\int_{-\infty}^{+\infty}\!\frac{d\omega}{2\pi}\sum_{n=-\infty}^{+\infty}\,e^{-i\left(\omega+n\Omega\right)t}e^{-i\omega t'}\,\mathcal{K}^{\alpha\beta}_n(\omega)\,.
	\end{equation}
	The period-averaged electromagnetic kernel $\mathcal{K}^{\alpha\beta}(\omega)$ is given by $\mathcal{K}_{n=0}^{\alpha\beta}(\omega)$. It reads
	\begin{equation}\label{eq: EM Kernel Floquet}
		\begin{split}
			\mathcal{K}_n^{\alpha\beta}(\omega)=2\sum_{m=-\infty}^{+\infty}\bar{b}_{\bQ,m} \bar{b}_{-\bQ,n-m+\eta}\delta_{\alpha\beta}-&i\int_\bq 4q_\alpha q_\beta\int_{-\infty}^{+\infty}\!\frac{d\nu}{2\pi}\,\sum_{m=-\infty}^{+\infty}\!\!\left[\boldsymbol{G}^R_{0,m}(\nu)\boldsymbol{G}^K_{m,n}(\nu-\omega)+\boldsymbol{G}^K_{0,m}(\nu)\boldsymbol{G}^A_{m,n}(\nu-\omega)\right]\\
			-&2i\int_\bq\int_{-\infty}^{+\infty}\!\frac{d\nu}{2\pi}\,\boldsymbol{G}^K_{0,n}(\nu)\,\delta_{\alpha\beta}\,,
		\end{split}
	\end{equation}
	where $\eta=0$ for period-preserving solutions and $\eta=1$ for period doubling ones.
	Integrating the last term by parts, 
	\change{
		we get
		\begin{equation}\label{eq: diama term by parts}
			\int_\bq\int_{-\infty}^{+\infty}\!\frac{d\nu}{2\pi}\,\boldsymbol{G}^K_{0,n}(\nu)\,\delta_{\alpha\beta}=-\int_\bq\int_{-\infty}^{+\infty}\!\frac{d\nu}{2\pi}\,\partial_{q_\alpha}\boldsymbol{G}^K_{0,n}(\nu)\,q_{\beta}\,.
		\end{equation}
		Using Eqs.~\eqref{eq: GRA Floquet} and \eqref{eq: GK Floquet}, one obtains 
		\begin{equation}\label{eq: dq GK}
			\partial_{q_\alpha}\boldsymbol{G}^K(\nu)=(2q_\alpha)\sum_{m=-\infty}^{+\infty}\left[\boldsymbol{G}^R_{0,m}(\nu)\boldsymbol{G}^K_{m,n}(\nu)+\boldsymbol{G}^K_{0,m}(\nu)\boldsymbol{G}^A_{m,n}\right]\,.
		\end{equation}
		Plugging~\eqref{eq: diama term by parts} and~\eqref{eq: dq GK} into Eq.~\eqref{eq: EM Kernel Floquet}, we obtain} the simple formula $\mathcal{K}_n^{\alpha\beta}(\omega\to0)=2\sum_{m=-\infty}^{+\infty}\bar{b}_{\bQ,m} \bar{b}_{-\bQ,n-m+\eta}\,\delta_{\alpha\beta}$. Fourier transforming to real time and real space, we get a space-time-dependent London penetration depth given by
	\begin{equation}
		[\lambda_L(\mathbf{x},t)]^{-2}\propto |\bar{b}(\mathbf{x},t)|^2\,.
	\end{equation}
	The optical conductivity can be obtained from the kernel as 
	\begin{equation}
		\sigma^{\alpha\beta}_n(\omega)=\frac{\mathcal{K}_n^{\alpha\beta}(\omega)}{i\omega}\,.
	\end{equation}
\end{widetext}

\end{document}